\documentclass[a4wide,11pt]{article}

\pdfoutput=1 
\usepackage[table]{xcolor}
\usepackage{colortbl}
\RequirePackage{ifpdf} 
\usepackage{amsmath,amsfonts} 
\usepackage{mathtools}
\usepackage{cases}
\usepackage[T1]{fontenc}
\usepackage{upquote,multirow,slashed}
\usepackage{authblk}
\usepackage{cite,appendix}
\usepackage{empheq}
\usepackage{lipsum}

\usepackage{jheppub}

\usepackage{pstricks}
\usepackage[final]{pdfpages} 
\usepackage{ifpdf} 
\usepackage{slashed}
\usepackage[normalem]{ulem}
\usepackage{color}
\usepackage{xcolor}
\definecolor{urlblue}{rgb}{0.2,0.4,0.7}
\definecolor{citegreen}{rgb}{0,0.6,0.2}
\definecolor{linkred}{rgb}{0.9,0.2,0.1}
\usepackage{hyperref}
\hypersetup{colorlinks=true, citecolor=citegreen, linkcolor=blue, urlcolor=blue}

\usepackage{graphics}
\usepackage{etoolbox} % Load before 'breqn' package to make 'lineno' work
\usepackage{fixmath}
\usepackage{psfrag}
\usepackage{autobreak}
\usepackage{marginnote}
\usepackage{enumitem}
\usepackage{appendix}
\usepackage{caption} 
\usepackage{notoccite} % If you have \cite commands in \section-like commands, or in \caption, the
\newcommand{\NOdisplay}[1]{ }

\def\MSbar{\overline{\mathrm{MS}}}
\def\TR{{\displaystyle \mathrm{T}_{F}}}

\def\gFF{\gamma_{{\scriptscriptstyle F\tilde{F}}}}
\def\g5{\gamma_5}
\def\oVFF{G_{5\mathrm{f\bar{f}}}}

\def\form{{\fontfamily{qcr}\selectfont
FORM}}

\title{An observation on Feynman diagrams with axial anomalous subgraphs in dimensional regularization with an anticommuting $\gamma_5$}

\author{Long Chen}
\emailAdd{longchen@sdu.edu.cn}
\affiliation{School of Physics, Shandong University, Jinan, Shandong 250100, China}

\abstract{
Through the calculation of the matrix element of the singlet axial-current operator between the vacuum and a pair of gluons in dimensional regularization with an anticommuting $\gamma_5$ defined in a Kreimer-scheme variant, we find that additional renormalization counter-terms proportional to the Chern-Simons current operator are needed starting from $\mathcal{O}(\alpha_s^2)$ in QCD. 
This is in contrast to the well-known purely multiplicative renormalization of the singlet axial-current operator defined with a non-anticommuting $\gamma_5$.
Consequently, without introducing compensation terms in the form of additional renormalization, the Adler-Bell-Jackiw anomaly equation does not hold automatically in the bare form in this kind of schemes.
We determine the corresponding (gauge-dependent) coefficient to $\mathcal{O}(\alpha_s^3)$ in QCD, using a variant of the original Kreimer prescription which is implemented in our computation in terms of the standard cyclic trace together with a constructively-defined $\gamma_5$.
Owing to the factorized form of these divergences, intimately related to the axial anomaly, we further performed a check, using concrete examples, that with $\gamma_5$ treated in this way, the axial-current operator needs no more additional renormalization in dimensional regularization but \textit{only} for non-anomalous amplitudes in a perturbatively renormalizable theory.
To be complete, we provide a few additional ingredients needed for a proposed extension of the algorithmic procedure formulated in the above analysis to potential applications to a renormalizable anomaly-free chiral gauge theory, i.e.~the electroweak theory.
}

\begin{document}

\allowdisplaybreaks[4]
\unitlength1cm
\maketitle
\flushbottom

\section{Introduction}
\label{sec:intro}

Dimensional regularization (DR)~\cite{tHooft:1972tcz,Bollini:1972ui} is the regularization framework that underlies most of the modern high-order perturbative calculations in the Standard Model (SM), especially in Quantum Chromodynamics (QCD). 
This is not only because the dimensionally regularized amplitudes are more neat compared to their counterparts defined in other regularization schemes, owing to many crucial symmetries well preserved in DR, but also because there are many powerful methods developed for evaluating Feynman loop integrals defined in DR. 
However, not all symmetries and algebraic operations in quantum field theories allow a clear and straightforward continuation from the 4- to D-dimensional spacetime, and in particular the intrinsically 4-dimensional object $\gamma_5$ defies such a continuous extension.
As well-known in literature, at the root of the issue is the contradiction between a fully anticommuting $\gamma_5$ in D ($\neq 4$) dimensions and the non-vanishing value of the cyclic trace of the products of a $\gamma_5$ and four $\gamma$ matrices in 4 dimensions.   
Nevertheless, the anticommutativity of $\gamma_5$ is essential for the concept of chirality of spinors in 4 dimensions.
On the other hand, a naive use of an anticommuting $\gamma_5$ in DR, where the invariance of loop integrals under arbitrary loop-momentum shifts is ensured, leads to the absence of the axial or Adler-Bell-Jackiw (ABJ) anomaly~\cite{Adler:1969gk,Bell:1969ts}.
In order to overcome these technical issues, various practical $\gamma_5$ prescriptions in DR have been developed in literature, e.g.~refs.~\cite{tHooft:1972tcz,Akyeampong:1973xi,Breitenlohner:1977hr,Bardeen:1972vi,Chanowitz:1979zu,Gottlieb:1979ix,Siegel:1979wq,Fujii:1980yt,Ovrut:1981ne,Espriu:1982bw,Buras:1989xd,Kreimer:1989ke,Korner:1991sx,Kreimer:1993bh,Larin:1991tj,Larin:1993tq,Chetyrkin:1997gb,Jegerlehner:2000dz,Ma:2005md,Tsai:2009it,Mihaila:2012pz,Fazio:2014xea,Moch:2015usa,Porto:2017asd,Bruque:2018bmy,Gnendiger:2017rfh,Zerf:2019ynn,Cherchiglia:2021uce,Rosado:2023ist}, ever since the invention of DR.

%These $\gamma_5$ prescriptions in DR can be broadly classified into two categories. 
Among these $\g5$ prescriptions, the original scheme by t' Hooft and Veltman~\cite{tHooft:1972tcz} (HV) as well as its variants~\cite{Akyeampong:1973xi,Breitenlohner:1977hr,Larin:1991tj,Larin:1993tq} is a very popular choice in the calculation of loop corrections, where the full anticommutativity of $\g5$ is sacrificed at least for bare amplitudes. 
The most celebrated property of these schemes based on a constructively-defined non-anticommuting $\g5$ is that the expression for any $\g5$-dependent object is mathematically unambiguously defined\footnote{In case the Lorentz indices of the Levi-Civita tensors introduced via the constructive definition of $\g5$ are not explicitly set 4-dimensional, one may need to impose an additional rule to fix the contraction ordering among multiple Levi-Civita tensors (see e.g.~refs~\cite{Moch:2015usa,Chen:2019wyb}), due to the lack of the 4-dimensional Schouten identity, at least when computing various divergent bare quantities individually. Because of this, we note that an unambiguous application of the variant in~\cite{Larin:1991tj,Larin:1993tq} is currently limited to pure QCD corrections. More comments on this are given later at the end of section~\ref{sec:RoK} and also the Appendix.}, in particular independent of the diagrams or matrix elements in which it may be embedded.
Apart from the apparent computational disadvantage when applied to the cases with multiple $\g5$ on a single fermion chain, another consequence of the loss of anticommutativity is that the Ward-Takahashi identities, %symmetry relations resulting from the local or extended gauge symmetries 
or their generalization, the Slanov-Taylor identities in non-Abelian gauge theories in Becchi-Rouet-Stora-Tyutin quantization~\cite{Becchi:1974md,Becchi:1975nq,Tyutin:1975qk}, are not respected at the level of bare amplitudes in chiral gauge theories determined in this kind of schemes: additional spurious anomalous terms appear in the bare expressions of dimensionally regularized $\g5$-dependent diagrams with ultraviolet (UV) divergences~\cite{Bardeen:1972vi,Chanowitz:1979zu,Gottlieb:1979ix,Trueman:1979en,Fujii:1980yt,Espriu:1982bw,Larin:1991tj,Larin:1993tq,Bos:1992nd}. 
However, owing to the locality of UV divergences, best summarized up in the famous Bogoliubov-Parasiuk-Hepp-Zimmermann (BPHZ) theorem~\cite{Itzykson:1980rh}, it is understandable that concerning perturbative corrections to Green functions with axial-current operators in a typical non-chiral gauge theory, e.g.~the QCD, the needed compensation terms can be traced back to just the overall UV divergence in the fermion-axial-coupling vertex corrections. 
Consequently, the corresponding (additional) renormalization constants can be systematically determined by requiring validity of relevant Ward-Takahashi identities, %up to at least the $\epsilon^0$-order for the finite remainders
and are valid to \textit{any} order in perturbation theory. 
%In this sense, the HV scheme and its variants based on a non-anticommuting $\g5$ is said to be applicable to all orders in perturbation theory.
In particular, this means that the results for the renormalization constants~\cite{Larin:1991tj,Larin:1993tq,Larin:1997qq,Rittinger:2012the,Ahmed:2021spj,Chen:2021gxv,Chen:2022lun} up to five-loop order in QCD for flavor non-singlet and singlet axial-current operators are universal, and should be applicable to any Feynman amplitudes or Green functions involving one or multiple external axial-current operators as long as one considers only their loop corrections generated by QCD interaction.\footnote{Note that regarding the subtle issue observed in $q\bar{q} \rightarrow ZH$ calculated with a non-anticommuting $\gamma_5$ reported in ref.~\cite{Ahmed:2020kme}, the loop corrections are generated using a heavy-top effective Lagrangian with the Higgs-gluon vertex. %One observes no such kind of subtlety in the perturbative correction to the matrix elements of the time-ordered product of an axial-current operator and a scalar operator generated by the standard QCD Lagrangian.
}
Application of a non-anticommuting $\gamma_5$ to the radiative corrections in the SM requires new counter-terms with new renormalization constants, which are discussed in refs.~\cite{Shao:2011tg,Belusca-Maito:2020ala,Belusca-Maito:2021lnk,Cornella:2022hkc}%up to one-loop order
, see e.g.~ref.~\cite{Belusca-Maito:2023wah} for a recent review of developments in this direction.

An alternative paradigm is to carefully define or manipulate the expressions of Dirac traces with $\g5$ keeping the full anticommutativity at least formally to certain extent, which has been pursued notably in refs.~\cite{Bardeen:1972vi,Chanowitz:1979zu,Gottlieb:1979ix,Buras:1989xd,Kreimer:1989ke,Korner:1991sx,Kreimer:1993bh,Chetyrkin:1997gb,Jegerlehner:2000dz,Mihaila:2012pz,Zerf:2019ynn}.
The conceptual advantage with the approaches based on a formally anticommuting $\g5$ over the previous ones is mainly that the need of the aforementioned non-trivial additional renormalizations for axial currents with, a priori, unknown renormalization constants, may be avoided from the outset.
The prescription developed in refs.~\cite{Kreimer:1989ke,Korner:1991sx,Kreimer:1993bh}, sometimes referred to as KKS scheme in literature, is particularly interesting, because regardless of whether the fermion chain with $\g5$ is open or closed in Feynman diagrams, $\g5$ is treated according to one systematic set of rules where the cyclicity of the Dirac traces with an odd number of $\g5$ is given up in favor of the formal anticommutativity of $\g5$, introducing the notion of non-cyclic $\g5$-trace with the so-called ``reading-point''.
To avoid confusion, let us emphasize that in the remainder of this article, by ``(original) Kreimer scheme'', we refer to the particular prescription formulated by Kreimer in his unpublished work~\cite{Kreimer:1993bh}, which contains a critical revision compared to the earlier discussions~\cite{Kreimer:1989ke,Korner:1991sx} as well as more specific statements regarding the treatment of axial anomalous graphs, the subject of the present publication.
Given the fact that a local gauge theory with an internal axial anomaly suffers from severe theoretical problems, e.g.~the loss of unitarity and all-order (multiplicative) renormalizability, we naturally limit ourselves to calculation of Feynman diagrams with at most an \textit{external} axial anomaly\footnote{An \textit{external} axial anomaly refers to the anomaly in the divergence of an external axial-current operator which is not coupled to the quantized gauge bosons in the Lagrangian of the theory. %Its appearance in the amplitudes or matrix elements or Green functions in question is merely due to insertion as an external local-composite operator, rather than being generated by the Lagrangian.
Axial anomalies of this kind are not only allowed in a gauge theory, but also crucial in understanding some important physical observations, such as a low-energy theorem for $\pi^0 \to \gamma \gamma$ decay~\cite{Sutherland:1967vf,Veltman:1967,Adler:1969gk,Bell:1969ts} and the solution of the so-called $U(1)$ problem~\cite{Weinberg:1975ui,tHooft:1976snw,tHooft:1986ooh}.} using Kreimer scheme in this work.

Reading prescriptions alternative to ref.~\cite{Kreimer:1993bh}, such as those allowed in the earlier refs.~\cite{Kreimer:1989ke,Korner:1991sx}, may lead to different results %(otherwise there would not be the need to revise at all)
for a closed fermion chain that agree with the latter only to the leading power in the Laurent expansion in the dimensional regulator $\epsilon$.\footnote{A recent non-trivial calculation of pure electroweak corrections at two-loop using the reading-point prescription of ref.~\cite{Korner:1991sx} can be found in ref.~\cite{Chen:2022mre}, and see also ref.~\cite{Heller:2020owb} for application to two-loop four-point mixed electroweak-QCD corrections.}
However, in case these closed fermion chains are embedded in Feynman diagrams with non-negative superficial UV divergences, the aforementioned difference can affect $\epsilon^0$ order or even pole terms for these diagrams at sufficiently high loop orders.  
It is not very obvious why in this kind of scenario Kreimer scheme~\cite{Kreimer:1993bh} can still work smoothly without any problem, as in general being unique is not yet sufficient to be correct.
While, in the highly non-trivial calculation of the four-loop strong-coupling $\beta$-function in the SM~\cite{Zoller:2015tha,Bednyakov:2015ooa} (in the limit of vanishing electroweak gauge-couplings), it turns out that the correct result~\cite{Poole:2019txl,Davies:2019onf} can follow from the reading prescription C in ref.~\cite{Bednyakov:2015ooa}, which is actually compatible with Kreimer scheme.
~\\

What motivated the present work is the following question: does the ABJ equation~\cite{Adler:1969gk,Bell:1969ts}, and the Adler-Bardeen theorem~\cite{Adler:1969er}, hold \textit{automatically} in the bare form, in the context of QCD corrections, with $\g5$ treated in Kreimer scheme~\cite{Kreimer:1993bh}?
Note that this is not the same question as whether an external flavor-singlet axial-current operator requires any UV renormalization: it does and this is well-known since Adler's seminal work~\cite{Adler:1969gk}. 
Hence, what the above question really refers to is rather whether the UV counter-term needed for an external singlet axial-current operator happens to cancel against the mixing term appearing in the renormalization of the axial anomaly operator with $\g5$ treated in this particular scheme.  
The sketch in the original ref.~\cite{Kreimer:1993bh} seems to indicate so, and this also seems to be the common lore as far as the author's impression is concerned due to $\g5$ being formally anticommuting in this scheme.
To our surprise, through the calculation of the matrix element of an external singlet axial-current operator between the vacuum and a pair of gluons, i.e.~the vector--vector--axial-vector (VVA) current-correlator, in QCD with an anticommuting $\gamma_5$ treated according to Kreimer scheme~\cite{Kreimer:1993bh},\footnote{apart from a different practical treatment of the Levi-Civita tensor, which we think should not affect the conclusion and will be discussed in detail later in the text} we find that the answer to the above question is, unfortunately, negative. 
More specifically, we find that additional renormalization counter-terms proportional to the Chern-Simons current operator are needed starting from $\mathcal{O}(\alpha_s^2)$ in QCD, in contrast to the well-known strictly multiplicative renormalization for this operator defined with a non-anticommuting $\gamma_5$.
Consequently, without introducing compensation terms by hand in the form of additional renormalization, the ABJ equation does not hold automatically in the bare form in this kind of schemes.
To compare, we checked explicitly that a parallel treatment with the axial-current vertex replaced by a pseudo-scalar vertex in topologically the same diagrams (with massive quark propagators) exhibits no such kind of issue at all, in the sense of resulting in the same 4-dimensional limit of finite remainders as those determined using a non-anticommuting $\gamma_5$. 
~\\

The remainder of this article is organized as follows.
In the next section~\ref{sec:RoK} we first recapitulate the essentials underlying the validity of the original Kreimer scheme, and explain how its variant is implemented in our computational set-up.
In section~\ref{sec:Calc} we expose the details of our calculations and results of the matrix elements of an external singlet axial-current operator and pseudo-scalar operator between the vacuum and a pair of external gluons in Kreimer scheme, at two different kinematical configurations, in QCD with massless and massive quarks. 
In particular, after providing the preliminaries in subsection~\ref{sec:prel}, we introduce the definitions of these quantities in subsection~\ref{sec:Calc-offshell}.  
In the third subsection~\ref{sec:obs} we present the main result of this work, the observation of an issue encountered in the aforementioned calculations, and try to discuss the implications.
We discuss briefly in the last subsection~\ref{sec:Calc-onshell} a few checks related to the aforementioned observation in order to clarify a few obvious questions. 
We conclude in section~\ref{sec:Conc}.

\section{Reformulation of Kreimer scheme using cyclic Dirac traces with $\hat{\g5}$}
\label{sec:RoK}

In this section, we first briefly review and then reformulate the essentials underlying Kreimer scheme in the unpublished work~\cite{Kreimer:1993bh}, albeit in terms of the standard Dirac trace in D dimensions with a constructively-defined non-anticommuting $\g5$, as this explains how this scheme is implemented in our technical set-up.
We hope this exposition could also be helpful to some practitioners who wish to give this scheme a try but, on the other hand, would like to utilize the familiar tools, e.g.~\form~\cite{Vermaseren:2000nd} as much as possible (where there is no built-in non-cyclic trace). 
Moreover, the procedure formulated in this recapitulation serves as the core routine in a  proposed prescription, in the Appendix~\ref{append:g5pre}, to deal with the non-anomalous amplitudes with $\g5$ in a renormalizable anomaly-free chiral gauge theory like the electroweak sector of the SM.
~\\

The validity of the original Kreimer scheme can be appreciated as follows. 
In the case of $\g5$ on open fermion lines, and the case with an even number of $\g5$ on the same fermion chain (closed or not), the applicability of an anticommuting $\g5$ is clear and long known in literature~\cite{Bardeen:1972vi,Chanowitz:1979zu,Buras:1989xd,Kreimer:1989ke}.
Essentially, after shifting $\g5$ anticommutatively along the fermion chain into some external structure, either a spinor or projector (including the case of a tree-level diagram), one ends up with a normal spinor-stripped, possibly off-shell, tensor amplitude for which the usual $\g5$-free renormalization procedure shall work: 
all the formal algebraic properties needed in the formal derivation of the symmetry-related structural relations of the theory, such as the Ward-Takahashi identities, are granted without encountering the algebraic inconsistency in taking $\g5$-odd traces in DR as alluded at the beginning of the Introduction~\ref{sec:intro};
And correspondingly, the correct results shall follow, without need to call for any additional renormalization pertaining to the $\g5$-vertices. 
After obtaining a UV-finite $\g5$-free tensor amplitude, one has the option to either pull back the $\g5$ safely or just leave it in the external structure, and the absence of $\g5$ issue now becomes evident in this treatment.
%No Dirac-trace with an odd number of $\g5$ in DR, no problem.
In principle, the possible appearance of intermediate infrared-soft (IR) divergences in the amplitudes with on-shell final states should not pose any conceptually new difficulties and does not require any extra compensation terms, provided that the very same treatment of $\g5$ and IR divergences is implemented consistently in the determination of the corresponding IR-subtraction counter-terms.

Extension to the closed fermion chain with an odd number of $\g5$ requires some additional insights, due to the well-known algebraic issue in taking $\g5$-odd traces in DR.
In case of a single $\g5$-odd fermion loop with negative superficial degree of UV divergence in a multiplicatively renormalizable theory, owing to the locality of UV divergences in Feynman diagrams summarized up in BPHZ theorem, all its UV divergences are sub-divergences; 
and those relevant to the renormalizations of $\g5$-vertices shall be traced back to the overall divergences of the aforementioned type of subgraphs with $\g5$ on open fermion lines.
We already know how to treat this kind of cases from the above, avoiding the appearance of spurious anomalous terms by anticommutatively shifting $\g5$ outside the non-singlet-type loop corrections to the fermion-axial-coupling vertex. 
Indeed, this treatment shall be adopted in order that the same renormalization constants determined using the previous kind of diagrams with open fermion lines~\cite{Bardeen:1972vi,Chanowitz:1979zu,Gottlieb:1979ix} can be employed here for diagrams with the $\g5$-odd fermion loop. (The exact positioning are determined by an algorithmic procedure to be presented below).
In a typical renormalizable gauge theory, there can be individual $\g5$-odd fermion loops that have non-negative superficial degree of UV divergence.  
However, as demonstrated systematically in ref.~\cite{Kreimer:1993bh}, a suitable coherent combination of all contributing loops attached with the same boson propagators (albeit, in different orderings) has, as a whole, an effective negative superficial UV degree, to which the aforementioned statement applies again.
From the technical side, this necessarily requires a unique unambiguous prescription to consistently evaluate all individual $\g5$-odd traces appearing in this set of contributing fermion loops, because there are intermediate spurious poles of possibly higher orders canceling between each other in the sum while different trace expressions can result from different positions of $\g5$ in traces in DR.
Based on the above insights, a systematic $\g5$ prescription is presented in ref.~\cite{Kreimer:1993bh} with the goal of not introducing any spurious anomalous terms that may violate gauge-symmetry relations, e.g.~Ward-Takahashi identities, in a renormalizable gauge theory. %This excludes gauge theories with internal gauge axial anomalies.  
We note that this formulation also takes into account additional conditions such as the charge-conjugation properties (Furry's theorem extended to include axial currents) which were not given adequate attention in earlier works~\cite{Kreimer:1989ke,Korner:1991sx}.%, very important to ensure correct evaluations of multi-loop amplitudes with axial currents.
~\\

Based on the above exposition of the underlying essentials, one may then appreciate that Kreimer scheme~\cite{Kreimer:1993bh} can also be implemented effectively in the following way, \textit{without} making any reference to the notion of the fancy non-cyclic $\g5$-trace introduced in the original refs.~\cite{Kreimer:1989ke,Korner:1991sx,Kreimer:1993bh}. % a concept that may intimidate some beginning practitioners.
Rather than touching the cyclicity of the trace (no trace employed in this work is non-cylic), the essence from our point of view is just that the relative position of a constructively-defined non-anticommuting $\g5$ in a Feynman diagram can not be arbitrary but shall obey certain rules indicated by the well-known treatment of the case of $\g5$ on open fermion lines~\cite{Bardeen:1972vi,Chanowitz:1979zu,Gottlieb:1979ix}.
In the bulk of this article, we focus on the QCD corrections to Green functions with  \textit{external} axial-current and/or pseudoscalar operators, but will provide in the Appendix~\ref{append:g5pre} the additional ingredients needed to be incorporated on top of the algorithmic procedure formulated below in order to apply to the anomaly-free electroweak theory.

Let us consider first a Feynman diagram $G$, used to denote also the corresponding Feynman integral, for the time-being only with one \textit{external} (axial-current) vertex $A_5$ with a $\g5$ on a %closed
fermion chain $F_c$. %(used to denote also the corresponding expression as a product of Dirac matrices).
The $F_c$ is written out in the direction \textit{against} the fermion-charge flow as in the standard convention, but is otherwise allowed to start from \textit{any} vertex or propagator chosen possibly by the diagram generator in use or the practitioner. %We stress that no trace employed in this work is non-cylic.
At this moment, the symbolic expression for $F_c$, as well as $G$, derived according to the Feynman rules of the theory in question should be regarded as merely a \textit{book-keeping} form. 
The mathematically unambiguous definition %still needs to be specified, and will be made ready only after the fermion chain $F_c$ with $\g5$ being defined according to 
is provided by an algorithmic manipulation of the input $F_c$ consisting of the following three main steps. %set of instructions/procedures/rules
\begin{figure}[htbp]
\begin{center}
\includegraphics[scale=0.88]{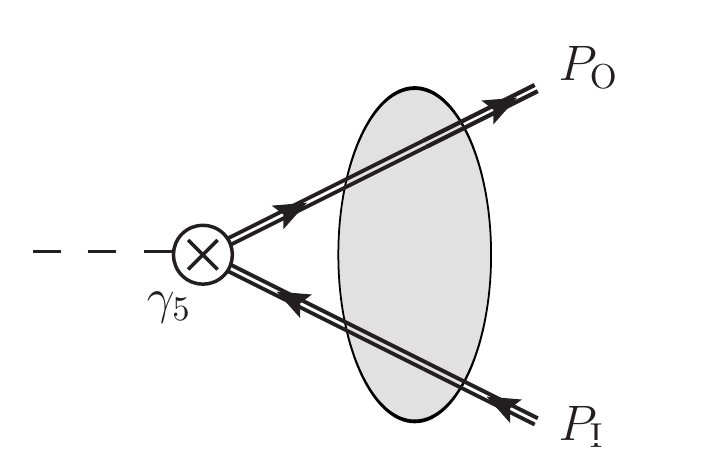}
\caption{An illustration of the I/O-leg identified for the maximal 1PI non-singlet-type fermion-axial-coupling subgraph $\oVFF$ where the $\g5$ matrix from an \textit{external} axial or pseudo-scalar coupling vertex (denoted by the circle with a cross) lies on this open continuous fermion chain indicated by the double line.
The arrows on the double lines represent the direction of the fermion-charge flow, which enters through the I-leg fermion propagator with incoming momentum $P_{\mathrm{I}}$ and leaves out via the  O-leg  fermion propagator with outgoing momentum $P_{\mathrm{O}}$, neither broken nor closed. The grey blob denotes the maximal 1PI non-singlet-type loop correction to the three-point fermion-axial-coupling vertex.
}
\label{fig:Max1PIopenVFF}
\end{center}
\end{figure}
\begin{itemize}
\item 
Step-1: 
Based on the graph information of $G$, identify the two fermion legs of the three-point fermion-axial-coupling subgraph $\oVFF$ with an external $A_5$ that contains the maximal one-particle-irreducible (1PI) non-singlet-type loop correction, as illustrated in figure~\ref{fig:Max1PIopenVFF}. % which are just two specific propagators on $F_c$. 
The phrase ``non-singlet-type'' refers to the graphical feature that the continuous fermion chain with the $\g5$ matrix from $A_5$ is left open in $\oVFF$.

As illustrated in figure~\ref{fig:Max1PIopenVFF}, the fermion leg of this subgraph $\oVFF$ with incoming fermion-charge flow is marked as I-leg and the corresponding fermion propagator reads as $S_F^{\mathrm{I}}(P_{\mathrm{I}})$. 
Similarly, the other fermion leg of $\oVFF$ with outgoing fermion flow is marked as O-leg and the corresponding fermion propagator $S_F^{\mathrm{O}}(P_{\mathrm{O}})$.

In practice, $\oVFF$ can be determined by examining all possible \textit{two-fermion cuts} through the fermion chain $F_c$, under the condition $P_{\mathrm{O}} -  P_{\mathrm{I}}$ equal to the momentum insertion through $A_5$, and then selecting the one resulting the largest 1PI subgraph with $A_5$.
Note that limited to the three-point graph $\oVFF$, the 1PI condition simply implies the absence of tree-propagators \textit{inside} $\oVFF$ with momenta either $P_{\mathrm{I}}$ or $P_{\mathrm{O}}$, in other words, $\oVFF$ is free of self-energy corrections to the I/O-legs. 
%The meaning of being the ``maximal'' is self-evident, simply the one which have all others as its proper subset.

\item 
Step-2: 
The final expression $\bar{F}_c$ for the fermion chain $F_c$ is \textit{defined} as the following average:
\begin{equation}
\label{eq:FCdef}
\bar{F}_c \equiv \frac{1}{2}\, \Big(F^{A_5 \rightarrow \mathrm{Ih}}_c + F^{A_5 \rightarrow \mathrm{Ot}}_c \Big)
\end{equation}
where $F^{A_5 \rightarrow \mathrm{Ih}}_c$ denotes a definite string of Dirac matrices obtained from the original bookkeeping form $F_c$ by anticommutatively shifting $\gamma_5$ from the original vertex $A_5$ to the \textit{head} of the I-leg propagator $S_F^{\mathrm{I}}(P_{\mathrm{I}})$ which is subsequently replaced by the following constructive expression~\cite{tHooft:1972tcz,Breitenlohner:1977hr}
\begin{equation}
\label{eq:gamma5}
	  -\frac{i}{4!}\epsilon^{\mu\nu\rho\sigma}\gamma_{\mu}\gamma_{\nu}\gamma_{\rho}\gamma_{\sigma} \equiv \hat{\g5} \,.
\end{equation}
The Lorentz indices of the Levi-Civita tensor $\epsilon^{\mu\nu\rho\sigma}$ in eq.~\eqref{eq:gamma5} should be 4-dimensional to be consistent with the relation $\g5^2 = \hat{1}$.\footnote{We use the convention $\epsilon^{0123} = -\epsilon_{0123} = +1$.}
To avoid confusion in notations, a new $\hat{\g5}$ with hat is introduced on the r.h.s.~of eq.~\eqref{eq:gamma5} to denote specifically this fixed expression, a non-anticommuting $\gamma_5$ matrix as in HV-scheme variants~\cite{tHooft:1972tcz,Breitenlohner:1977hr}.
Similarly, $F^{A_5 \rightarrow \mathrm{Ot}}_c$ denotes a definite string of Dirac matrices obtained by inserting eq.~\eqref{eq:gamma5}, albeit at the \textit{tail} of the O-leg propagator $S_F^{\mathrm{O}}(P_{\mathrm{O}})$ in $F_c$ and taking into account the relative sign generated by anticommutatively shifting $\g5$ to this position from $A_5$.

In the case of a single \textit{external} $A_5$ on $F_c$, the above specification is sufficient to guarantee an unambiguous expression for $F^{A_5 \rightarrow \mathrm{Ih}}_c$ and $F^{A_5 \rightarrow \mathrm{Ot}}_c$ respectively, independent of any particular choice of the starting point from which $F_c$ is chosen to be written out (in the direction against the fermion-charge flow). %, because each is now a standard cyclic trace.
The average in eq.~\eqref{eq:FCdef} is necessary to ensure the validity of Furry's theorem extended to include axial currents at the bare level in this prescription~\cite{Kreimer:1993bh}. 

\item 
Step-3: 
After the Dirac algebra done according to the above prescription, %in D dimensions, 
the tensor loop integrals in $G$ are then defined and evaluated in conventional dimensional regularization~\cite{tHooft:1972tcz,Bollini:1972ui,Collins:1984xc} in the usual way. 
%Only now the definition of $G$ with one $A_5$ in dimensional regularization can be considered as \textit{completed}, and given by the unique expression resulting from the above procedure.

\end{itemize}

In case of an even number of $\g5$ on the same fermion chain $F_c$ in $G$, the relation $\g5^2 = \hat{1}$ can be applied, resulting in a unique trace expression free of $\g5$ for $F_c$.\footnote{The independence of the resulting trace for a closed fermion chain with an even number of $\g5$ on the choice of positions to which the pairs of $\g5$ are anticommuted and annihilated using $\g5^2 = \hat{1}$ can be appreciated in the following way. 
By the virtue of $\big(\slashed{p} + m \big)\, \gamma^{\mu} \, \g5 = \g5 \, \big(\slashed{p} -m \big)\, \gamma^{\mu} $, there will be, at most, two symbolically different expressions, related to each other by flipping the signs of all fermion-propagator masses on the closed fermion chain (in absence of Yukawa couplings).
%If the sign of the Yukawa coupling is correlated with the mass sign of the fermion involved, the same statement holds as well.
If one considers only gauge interactions as well as Yukawa couplings correlated with fermion masses in sign, such as those in the SM, then the terms odd under a homogeneous sign-flip of all fermion-propagator masses do not contribute due to the vanishing traces of an odd number of Dirac-$\gamma$ matrices.
Consequently, the two aforementioned expressions are algebraically equivalent for closed fermion chains with an even number of $\g5$ (regardless of being on-shell cut or not).} 
In case of an odd number $N_e$($\geq 3$) of external $A_5$ vertices appearing on $F_c$ in $G$, another level of average is needed to reach an unambiguous trace result for $F_c$.
To this end, after applying $\g5^2 = \hat{1}$ before entering the above algorithmic procedure, only one $\g5$ is left at certain original external vertex, denoted as $A^{[i]}_5$, and the subsequent application of the above procedure leads to a definite trace expression, given specifically by eq.~(\ref{eq:FCdef}) with eq.~(\ref{eq:gamma5}), denoted as $\bar{F}^{[i]}_c$.
Then the final unique expression $\bar{\bar{F}}_c$ for this fermion chain is defined as the following average 
\begin{equation}
\label{eq:FCdefG}
\bar{\bar{F}}_c \equiv \frac{1}{N_e}\, \sum_{i=1}^{N_e} \bar{F}^{[i]}_c \,,
\end{equation}
over the given set of $N_e$ \textit{external} $\g5$-vertices of this fermion loop.
We note that analogous to the case with an even number of $\g5$ on the same fermion chain, the position where any pair of $\g5$ is brought together anticommutatively is irrelevant for the resulting expression and can be chosen at will, but the application of $\g5^2 = \hat{1}$ shall be done at least \textit{before} inserting eq.~\eqref{eq:gamma5}.

Unlike in the original refs.~\cite{Korner:1991sx,Kreimer:1993bh}, we tend to regard the average (\ref{eq:FCdefG}) as a prescription to guarantee an unambiguous trace definition for each closed fermion chain with an odd $N_e$($\geq 3$) external $A_5$-vertices (all treated on the same foot), which a priori has nothing to do with the requirement of Bose symmetry at the level of amplitude.
%Namely we view eq.~(\ref{eq:FCdefG}) merely as a part of the internal proper definition of the trace within the very scope of one particular fermion chain (not several fermion chain as a whole), irrespective of the bosons coupled to the corresponding $\g5$-vertices; namely eq.~(\ref{eq:FCdefG}) is a chain-by-chain definition, rather at the scope of amplitude.
In particular, we demand the average~(\ref{eq:FCdefG}) for each single fermion chain, \textit{irrespective} of whether these external $A_5$ vertices are axial-current or pseudo-scalar operators.% associated with possibly identical gauge or scalar bosons.
\footnote{We prefer to have, in this computational scheme, a definite one-loop result for a triple-current correlator with generic scalar-, vector- and axial-component, whose 4-dimensional limit equals to that determined in HV scheme, irrespective of the identities of possibly coupled fields.} 
On the other hand, in case several $A_5$ vertices happen to couple to \textit{identical} gauge bosons, eq.~(\ref{eq:FCdefG}) is necessary to ensure the Bose symmetry for the resulting amplitude determined in Kreimer scheme.
%Of course, in case the mathematical formula for each individual trace term with $\g5$ would have been done using the standard 4-dimensional Dirac algebra, then one would have observed that each single trace term in eq.~(\ref{eq:FCdef}) and eq.~(\ref{eq:FCdefG}) would lead to the same mathematically-equivalent expression; consequently the effect of averaging in these two definitions reduces to an trivial identity: $1/N*(N*x) = x$. However, with the insertion of \eqref{eq:gamma5} at different positions which does not fully anticommute with generic Dirac matrices, whether this average is performed or not can lead to difference at the relatively $\epsilon$-suppressed order.
This completes the description of the algorithmic procedure we employed to define the trace of a fermion chain with $\g5$ in DR as far as QCD corrections to Green functions with given external axial-current and/or psedoscalar operators are concerned.
In this kind of situations, the set of $N_e$ \textit{external} $\g5$-vertices to be averaged over in eq.\eqref{eq:FCdefG} shall be clear from the outset.
Whereas, in order to extend the application of the above procedure to the anomaly-free electroweak theory, one needs to first determine explicitly the set of external vertices (or rather the corresponding external momenta) identified w.r.t.~the fermion loop in question before applying the above procedure, which are discussed in the Appendix~\ref{append:g5pre_IE} for readers with interest.

Therefore, from the pure technical point of view, we did not need to appeal to the fancy notion of a non-cyclic $\g5$-trace, as advocated in the original refs.~\cite{Kreimer:1989ke,Korner:1991sx,Kreimer:1993bh}.
All we need in our computational set-up employed for the calculations discussed in the next section are just the standard cyclic Dirac trace and the constructive expression~\eqref{eq:gamma5} for $\g5$, which will be inserted \textit{outside} the properly identified unique $\oVFF$ for each $A_5$ indicated by the well-known treatment of the open-fermion-line case~\cite{Bardeen:1972vi,Chanowitz:1979zu,Gottlieb:1979ix}, more specifically, at the head of $S_F^{I}(P_{\mathrm{I}})$ and tail of $S_F^{O}(P_{\mathrm{O}})$ in eq.~\eqref{eq:FCdef}~\cite{Kreimer:1993bh}. 
The average~(\ref{eq:FCdefG}) is not yet needed as far as our present calculation is concerned, where we have only one axial-current or pseudo-scalar operator. 
The evaluation of the resulting standard $\g5$-free cyclic Dirac traces %free of $\g5$ or Levi-Civita tensor 
in D dimensions, allowed to be started from any vertex or propagator, can be conveniently done by a computer-algebra tool such as \form.
~\\

A comment is in order regarding the treatment of the Levi-Civita tensor $\epsilon^{\mu\nu\rho\sigma}$, introduced e.g.~via eq.~\eqref{eq:gamma5} for $\gamma_5$, an aspect in which our later calculations differ slightly from the original Kreimer scheme. 
As well-known in literature, strictly speaking the Levi-Civita tensor $\epsilon^{\mu\nu\rho\sigma}$ can only be mathematically defined consistently in 4 dimensions. %, just like the absence of a fully-anticommuting $\g5$ in D($\neq 4$) dimensions.
To be more specific, the mathematical inconsistency appears once one insists on the commutation in the contraction ordering for a product of multiple $\epsilon^{\mu\nu\rho\sigma}$~\cite{Breitenlohner:1977hr,Siegel:1980qs}, due to the lack of the 4-dimensional Schouten identity.
However, %Just like the fact that one can indeed formulate a $\gamma_5$-prescription in DR where the anticommutativity is maintained at least formally to some extent, showcased by the very $\gamma_5$-scheme discussed in this article, 
the aforementioned mathematical inconsistency does not exclude, in principle, the possibility of carefully manipulating $\epsilon^{\mu\nu\rho\sigma}$ with D($\neq 4$)-dimensional Lorentz indices but without encountering any inconsistency for a particular problem.%, which are actually practiced in many successful perturbative calculations.

As far as the calculations involved in the present work are concerned, we have only a single external axial-current operator in Feynman diagrams, and never need to apply the relation $\g5^2 = \hat{1}$. Furthermore, we have just one pair of Levi-Civita tensors to be contracted, according to 
\begin{eqnarray} \label{eq:LeviCivitaContRule}
\epsilon^{\mu\nu\rho\sigma} \epsilon^{\mu'\nu'\rho'\sigma'} 
= \mathrm{Det}\Big[g^{\alpha \alpha'} \Big]~, 
\text{~  with $\alpha \in \{\mu,\nu,\rho,\sigma\}$ and $\alpha' \in \{\mu',\nu',\rho',\sigma'\}$,} 
\end{eqnarray}
and there is no ambiguity or need to impose a definite convention in the contraction ordering~\cite{Moch:2015usa,Chen:2019wyb}.
We thus conclude that no mathematical inconsistency will be generated by setting the dimensionality of the spacetime-metric tensors on the r.h.s.~of eq.~\eqref{eq:LeviCivitaContRule} to be $D\neq 4$ in our calculations in the next section. 
Alternatively, one may also appreciate this statement from the perspective of form factor decomposition for the $\g5$/$\epsilon^{\mu\nu\rho\sigma}$-free tensor amplitudes in the underlying D dimensions together with the fact that the $\MSbar$ UV renormalization can be performed at the tensor level leading to (UV) finite tensor amplitudes.
%More comments from this perspective will be given in the next section. Alternatively, one may refer to the discussion in ref.~\cite{Mondejar:2012sz}.

On the other hand, there is indeed certain technical convenience brought by such a treatment of $\epsilon^{\mu\nu\rho\sigma}$ as advocated in~refs.\cite{Larin:1991tj,Zijlstra:1992kj,Larin:1993tq} for our calculations.
We do not need to implement the dimensional splitting in the Lorentz algebra like in HV scheme, and moreover we are allowed to complete the contraction of $\epsilon^{\mu\nu\rho\sigma}$, and also the contraction of the Lorentz indices of resulting spacetime-metric tensors with those of the tensor amplitudes, before completing tensor loop integrals in conventional dimensional regularization~\cite{tHooft:1972tcz,Bollini:1972ui,Collins:1984xc}. 
This is because %, with all Lorentz indices uniformally D-dimensional, 
these operations now commute with each other, and this kind of treatment has been successfully practiced  a lot, e.g.~in refs.~\cite{Larin:1991tj,Zijlstra:1992kj,Larin:1993tq,Moch:2015usa,Blumlein:2021ryt,Chen:2019wyb}, without encountering any problems.
One may be allowed to do so in the application of a Kreimer-scheme variant only if the potential compatibility issue with $\g5^2 = \hat{1}$ and/or the contraction-order ambiguity of the Levi-Civita tensors are properly taken care of for the calculations at hand. %which can of course be avoided simply by taking the Levi-Civita tensor to be 4 dimensional.

As far as the pure QCD corrections to Green functions or matrix elements of time-ordered products of local-composite operators involving multiple $\g5$ are concerned, the main focus of this work to be presented below, it shall thus be clear that the above non-4-dimensional treatment of $\epsilon^{\mu\nu\rho\sigma}$ is not only feasible but also convenient to use.
In the Appendix~\ref{append:g5pre_LV}, we extend further the applicability of this kind of technically convenient non-4-dimensional treatment of $\epsilon^{\mu\nu\rho\sigma}$ by providing a prescription to fix the contraction-order ambiguity among multiple $\epsilon^{\mu\nu\rho\sigma}$,  %without the aforementioned compatibility issue with the treatment of $\g5$,
expected to work for a large variety of cases in practical calculations.% with direct phenomenological applications.

\section{Calculation of the VVA diagrams using Kreimer-scheme variants}
\label{sec:Calc}

We are now in position to expose the technical aspects of our calculations and discuss the results in Kreimer scheme~\cite{Kreimer:1993bh}, implemented according to the description in the previous section.

\subsection{Preliminaries}
\label{sec:prel}

For the sake of readers' convenience, we first briefly recapitulate the ABJ anomaly equation with a non-anticommuting $\g5$, along which conventions and notations to be used in later discussions are introduced too.
%We follow mainly the notations and conventions in ref.~\cite{Chen:2021gxv}.
The all-order ABJ anomaly equation~\cite{Adler:1969gk,Adler:1969er} for an external singlet axial current in QCD with $n_f$ massless quarks, expressed in terms of renormalized local-composite operators, reads: %
\begin{eqnarray} 
\label{eq:ABJanomalyEQ}
\big[\partial_{\mu} J^{\mu}_{5,s} \big]_{R} = a_s\, n_f\, \TR \,  \big[F \tilde{F} \big]_{R}\,,
\end{eqnarray}
where the subscript $R$ at a square bracket denotes operator renormalization, $a_s \equiv \frac{\alpha_s}{4 \pi} = \frac{g_s^2}{16 \pi^2}$ is a shorthand notation for the QCD coupling and $\TR=1/2$. 
%Note that this equation is meant to hold in 4 dimensions, for the 4-dimensional limit of the finite remainders of their matrix elements or Green functions where they are embedded. 
The exact meaning of the renormalization operation applied to the operators in both sides of the above equation depends, among others, on the $\g5$-schemes in DR.

The proper definition for the hermitian singlet axial current $J^{\mu}_{5,s}$ in eq.~(\ref{eq:ABJanomalyEQ}) in schemes based on a non-anticommuting $\hat{\g5}$ involves the following matrix~\cite{Akyeampong:1973xi,Fujii:1980yt,Collins:1984xc,Larin:1991tj,Larin:1993tq}:
\begin{align}
\label{eq:gamma5-axial}
\frac{1}{2}\big( \gamma^{\mu} \hat{\g5} - \hat{\g5} \gamma^{\mu}\big) 
= -\frac{i}{3!} \epsilon^{\mu\nu\rho\sigma} \gamma_{\nu} \gamma_{\rho} \gamma_{\sigma}\,,
\end{align}
with eq.~\eqref{eq:gamma5} inserted for $\hat{\g5}$ on the l.h.s.
Note that eq.~\eqref{eq:gamma5-axial} is an exact mathematical identity, following from the total antisymmetry of $\epsilon^{\mu\nu\rho\sigma}$ and the defining relations of the Dirac algebra, independent of the dimensionality of the Lorentz indices. 
As one can check, the ``symmetrization'' in the definition~\eqref{eq:gamma5-axial} for a hermitian axial-current operator is necessary to ensure that the Furry's theorem is respected exactly at the bare level despite the loss of full anticommutativity for $\hat{\g5}$.
To be specific, the singlet axial-current operator with a non-anticommuting $\g5$ is defined as $\big[ J^{\mu}_{5,s}\big]_{B} = \sum_{i=1}^{n_f} \, \bar{\psi}^{B}_{i}  \, \frac{1}{2}\big( \gamma^{\mu} \hat{\g5} - \hat{\g5} \gamma^{\mu}\big) \, \psi^{B}_i$. 
On the r.h.s.~of eq.(\ref{eq:ABJanomalyEQ}), $F \tilde{F} \equiv  - \epsilon^{\mu\nu\rho\sigma} F^a_{\mu\nu} F^a_{\rho\sigma} \equiv \epsilon_{\mu\nu\rho\sigma} F^a_{\mu\nu} F^a_{\rho\sigma}$ denotes the contraction of the field-strength tensor and its dual, with $F^a_{\mu\nu} \equiv \partial_{\mu} A_{\nu}^{a} - \partial_{\nu} A_{\mu}^{a} + g_s \,  f^{abc} A_{\mu}^{b} A_{\nu}^{c}$, $A_\mu^a$ the gluon field, and $f^{abc}$ the non-Abelian color group structure constants.

With eq.~\eqref{eq:gamma5-axial}, the renormalization of the operators $\partial_{\mu} J^{\mu}_{5,s}$ and $F \tilde{F}$ can be arranged into the following matrix form~\cite{Adler:1969gk,Trueman:1979en,Espriu:1982bw,Breitenlohner:1983pi}: %
\begin{eqnarray}
\label{eq:Zsmatrix}
\begin{pmatrix}
\big[\partial_{\mu} J^{\mu}_{5,s} \big]_{R}\\
\big[F \tilde{F}\big]_{R}
\end{pmatrix}
= 
\begin{pmatrix}
Z_{s} &  0 \\
Z_{FJ} &  Z_{F\tilde{F}}
\end{pmatrix}
\cdot 
\begin{pmatrix}
\big[\partial_{\mu} J^{\mu}_{5,s} \big]_{B}\\
\big[F \tilde{F}\big]_{B}
\end{pmatrix}\,,
\end{eqnarray}
where the subscript $B$ at a square bracket indicates that all fields therein are bare. 
Eq.~\eqref{eq:Zsmatrix} shows that the axial-current operator defined with a non-anticommuting $\hat{\g5}$ in eq.~\eqref{eq:gamma5-axial} renormalizes multiplicatively. 
However, the renormalization of the axial-anomaly operator $F\tilde{F}$, which can be written as the divergence of the gauge-non-invariant Chern-Simons current $K^{\mu}$ following from an mathematical identity,
\begin{eqnarray} 
\label{eq:Kcurrent}
F \tilde{F}  &=& \partial_{\mu} K^{\mu} \,=\, \partial_{\mu} \Big(-4 \,\epsilon^{\mu\nu\rho\sigma} \,\Big(A_{\nu}^{a} \partial_{\rho} A_{\sigma}^{a} \,+\, g_s\,\frac{1}{3} f^{abc} A_{\nu}^{a} A_{\rho}^{b} A_{\sigma}^{c} \Big) \Big)\,,
\end{eqnarray}
is not strictly multiplicative but involves mixing with the divergence of $J^{\mu}_{5,s}$~\cite{Adler:1969gk,Espriu:1982bw,Breitenlohner:1983pi,Luscher:2021bog}. 
In fact, the renormalization for the pair of scalar operators $\partial_{\mu} J^{\mu}_{5,s}$ and $F \tilde{F} $ are known to derive from that of the corresponding Lorentz currents $J^{\mu}_{5,s}$ and $K^{\mu}$~\cite{Espriu:1982bw,Breitenlohner:1983pi,Larin:1993tq,Luscher:2021bog}.
In other words, these two sets of operators share exactly the same renormalization constants introduced in eq.~\eqref{eq:Zsmatrix}.

A few comments are in order.
In the so-called Larin's prescription or scheme~\cite{Larin:1991tj,Larin:1993tq}, %Strictly speaking, Larin's scheme refers to the following convention in determining the renormalization constant for the singlet axial current operator  
$Z_{F\tilde{F}}$ and $Z_{FJ}$ are chosen to be in $\MSbar$ renormalization convention, while the multiplicative renormalization constant $Z_s$ of the singlet axial current determined by demanding eq.~\eqref{eq:ABJanomalyEQ} in 4 dimensional limit, is not pure $\MSbar$. 
In particular, $Z_{F\tilde{F}}$ in $\MSbar$ scheme is equal to the $\MSbar$ QCD-coupling renormalization constant $Z_{a_s}$, as verified explicitly to $\mathcal{O}(\alpha_s^4)$ in ref.~\cite{Ahmed:2021spj} and proven rigorously in ref.~\cite{Luscher:2021bog}. 
(The proof assumes a proper treatment of the fully antisymmetric structure involved in these composite operators, such as in~refs.\cite{Larin:1991tj,Larin:1993tq}, and does not have to carry over to other treatments of the $\g5$ in different regularization prescriptions.)
Consequently, $\gFF$ is equal to the negative of the QCD $\beta$-function defined as $\beta \equiv - \frac{\mathrm{d} \ln Z_{a_s}}{\mathrm{d} \ln \mu^2}$.
$Z_s$ is conveniently parameterized as the product of a pure $\MSbar$-renormalization factor $Z^{ms}_{s}$ and an additional finite, \textit{by definition} $\epsilon$-independent, renormalization factor $Z^{f}_{s}$, namely $Z_{s} \equiv Z^{f}_{s} \, Z^{ms}_{s}$.
This is because eq.~\eqref{eq:ABJanomalyEQ} is meant and required to hold only in the 4-dimensional limit at each perturbative order (after proper subtraction of possible IR divergences when sandwiched between on-shell external states).
The current most precise result for $Z_{s}$ in QCD is of $\mathcal{O}(\alpha_s^5)$ provided in ref.~\cite{Chen:2022lun}.
~\\

Provided that $\g5$ in $J^{\mu}_{A_5,s} \equiv \sum_{i=1}^{n_f} \, \bar{\psi}^{B}_{i}  \,\gamma^{\mu} \g5 \, \psi^{B}_i$ is treated in Kreimer scheme,  eq.~\eqref{eq:ABJanomalyEQ} would be \textit{expected}, as implied in ref.~\cite{Kreimer:1993bh}, to be equivalent to%
\begin{equation}
\label{eq:ABJNaiveBare}
\big[\partial_{\mu} J^{\mu}_{A_5,s} \big]_{B} = \hat{a}_s\, n_f\, \TR \,  \big[F \tilde{F} \big]_{B} \,,     
\end{equation}
holding automatically at the bare level (with a mass-dimensionless definition of the bare coupling constant $\hat{a}_s$), as in the original work~\cite{Adler:1969er} in Quantum Electrodynamics using Pauli-Villars regularization~\cite{Pauli:1949zm} with an anticommuting $\g5$ (see also ref.~\cite{Espriu:1982bw} using the hybrid $\g5$ prescription~\cite{Chanowitz:1979zu} up to $\mathcal{O}(a^2_s)$). %where the strictly 4-dimensional finite expression for the ABJ-triangle fermion loop is hard-coded in the calculation. 
Note that this expectation~\eqref{eq:ABJNaiveBare} does not necessarily imply that $\big[\partial_{\mu} J^{\mu}_{A_5,s} \big]_{B}$ equals to $\big[\partial_{\mu} J^{\mu}_{A_5,s} \big]_{R}$, which was long known to be incorrect due to the axial anomaly~\cite{Adler:1969gk}. 
Instead, the \textit{expectation} would read
\begin{eqnarray}
\label{eq:ABJNaiveRen}
\big[\partial_{\mu} J^{\mu}_{A_5,s} \big]_{R} =  Z^{A_5}_s \, \big[\partial_{\mu} J^{\mu}_{A_5,s} \big]_{B} \,, \quad \big[F \tilde{F}\big]_{R}  = Z_{F\tilde{F}} \, \big[F \tilde{F}\big]_{B} \,+\, Z_{KJ} \big[\partial_{\mu} J^{\mu}_{A_5,s} \big]_{B}\,,
\end{eqnarray}%
where anticipating the potential difference compared to the renormalization constants introduced in eq.~\eqref{eq:Zsmatrix} for a non-anticommuting $\g5$ scheme, two new notations $Z^{A_5}_s$ and $Z_{KJ}$ are introduced.
Demanding the ABJ equation in terms of the renormalized operators defined in eq.\eqref{eq:ABJNaiveRen} holding in the same form as eq.~\eqref{eq:ABJanomalyEQ}, then the UV counter-term needed for $\big[\partial_{\mu} J^{\mu}_{A_5,s} \big]_{B}$ on the l.h.s.~of eq.~\eqref{eq:ABJNaiveBare} determined in this anticommuting $\g5$ scheme is expected to cancel against the mixing term $Z_{KJ} \, \big[\partial_{\mu} J^{\mu}_{5,s} \big]_{B}$ appearing in the renormalization of $\big[F \tilde{F}\big]_{B}$ on the r.h.s.; after all divergences cancel one takes the 4-dimensional limit where the equality is expected.
More specially, combining eq.~\eqref{eq:ABJNaiveBare} and  eq.\eqref{eq:ABJNaiveRen} in this $\g5$ scheme one would then expect $Z_{F\tilde{F}} = Z_{a_s}$ (with ${\hat a}_s = Z_{a_s} \, a_s$) and $Z^{A_5}_s - a_s\, n_f\, \TR \, Z_{KJ} = 1$.
What we will show below in detail is that our calculations do not support this expectation using Kreimer scheme (or rather its variant).

\subsection{The anomalous form factor at zero momentum insertion}
\label{sec:Calc-offshell}

First we consider the matrix element of the singlet axial-current operator $J^{\mu}_{A_5,s}$ between the vacuum and a pair of off-shell gluons in QCD with $n_f$ massless quarks. 
We denote by $\langle 0| \hat{\mathrm{T}}\left[ J^{\mu}_{A_5,s}(y) \, A_a^{\mu_1}(x_1) \, A_a^{\mu_2}(x_2) \right] |0 \rangle |_{\mathrm{amp}}$ the amputated 1PI vacuum expectation value of the time-ordered covariant product of the (singlet) axial current and two gluon fields in coordinate space with open Lorentz indices.
Subsequently, we introduce the following rank-3 matrix element $\Gamma^{\mu \mu_1 \mu_2}_{lhs}(p_1, p_2)$ in momentum space, 
\begin{eqnarray}
\label{eq:Glhs1PI}
\Gamma^{\mu \mu_1 \mu_2}_{lhs}(p_1, p_2) \equiv 
\int d^4 x  d^4 y \, e^{-i p_1 \cdot x - i q \cdot y }\,  \langle 0| \hat{\mathrm{T}}\left[ J_{A_5,s}^{\mu}(y) \, A_a^{\mu_1}(x) \, A_a^{\mu_2}(0) \right] |0 \rangle |_{\mathrm{amp}} 
\end{eqnarray}
where translation invariance has been used to shift the coordinate of one gluon field to the origin. 
The resulting total momentum conservation factor, ensuring $p_2 = -q - p_1$ for the momentum flow through the gluon field $A_a^{\mu_2}(0)$, is left implicit.

\begin{figure}[htbp]
\begin{center}
\includegraphics[scale=0.45]{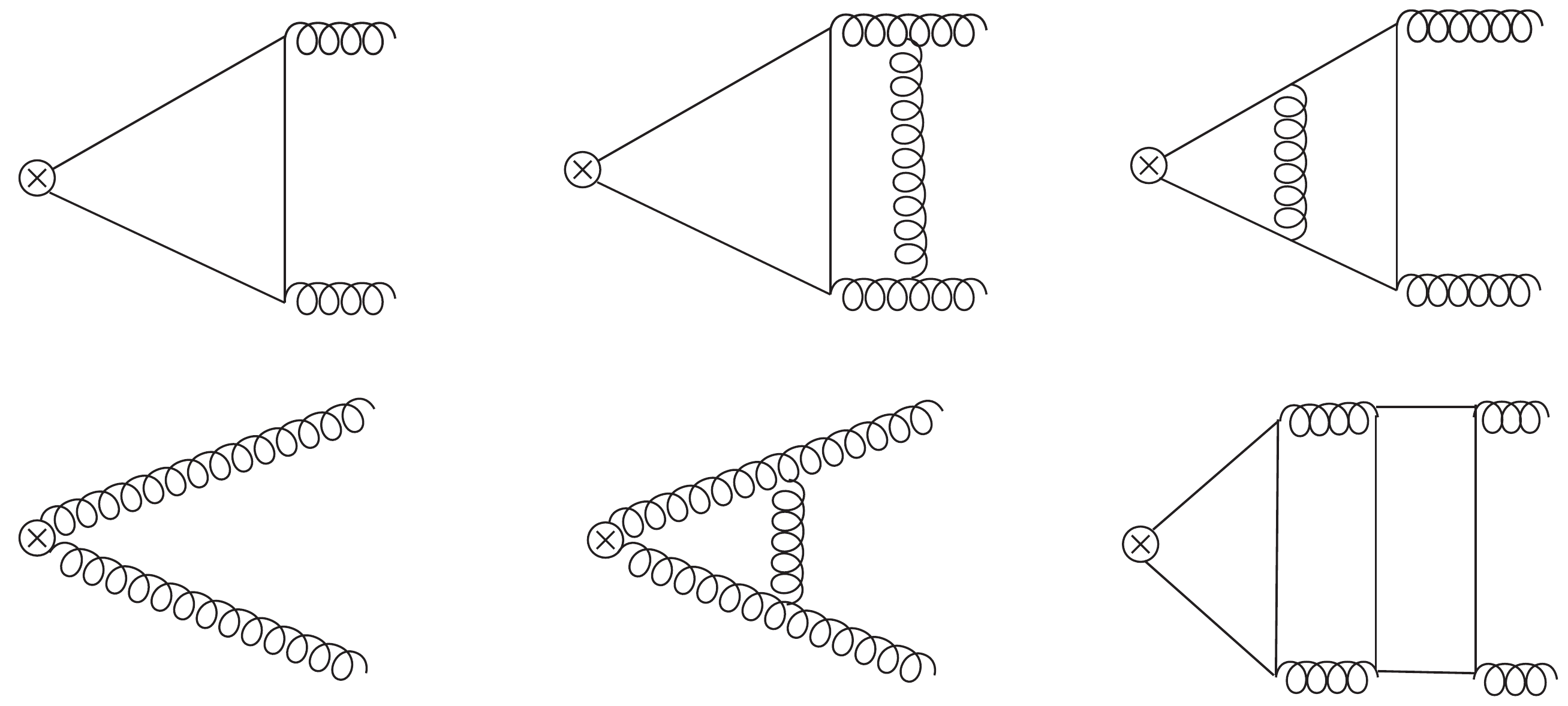}
\caption{Representative Feynman diagrams for the matrix elements of operators in 
the ABJ equation~\eqref{eq:ABJanomalyEQ} between vacuum and a pair of external gluons 
up to $\mathcal{O}(a_s^3)$. The axial-current operator is indicated by the circle with a cross inside, and the directions of fermion-charge flow on quark propagators are suppressed.
}
\label{fig:samplediagrams}
\end{center}
\end{figure}

At leading order, this is just the matrix element corresponding to the famous one-loop fermion-triangle diagram~\cite{Adler:1969gk,Bell:1969ts} with the polarization vectors of the external gluons stripped off and without imposing on-shell constraints on the incoming gluon momenta.
The higher order corrections consist of all 1PI diagrams with amputated external gluon legs to the specific loop order in question, which are known to be not vanishing~\cite{Adler:1969gk,Anselm:1989gi,Larin:1993tq,Mondejar:2012sz}, especially concerning the so-called ``light-by-light'' corrections starting from three-loop orders.{\color{blue}\footnote{Note that this was known~\cite{Adler:2004qt} to be not in contradiction with Adler-Bardeen theorem~\cite{Adler:1969er}.}}
Representative Feynman diagrams up to three-loop orders are given in figure~\ref{fig:samplediagrams}.

For the r.h.s.~of the ABJ equation~\eqref{eq:ABJanomalyEQ}, we define in analogy to eq.~\eqref{eq:Glhs1PI}:
\begin{eqnarray}
\label{eq:Grhs1PI}
\Gamma^{\mu \mu_1 \mu_2}_{rhs}(p_1, p_2) \equiv 
\int d^4 x  d^4 y \, e^{-i p_1 \cdot x - i q \cdot y }\,  \langle 0| \hat{\mathrm{T}}\left[ K^{\mu}(y) \, A_a^{\mu_1}(x) \, A_a^{\mu_2}(0) \right] |0 \rangle |_{\mathrm{amp}} \,.
\end{eqnarray}

\begin{figure}[htbp]
\begin{center}
\includegraphics[scale=0.5]{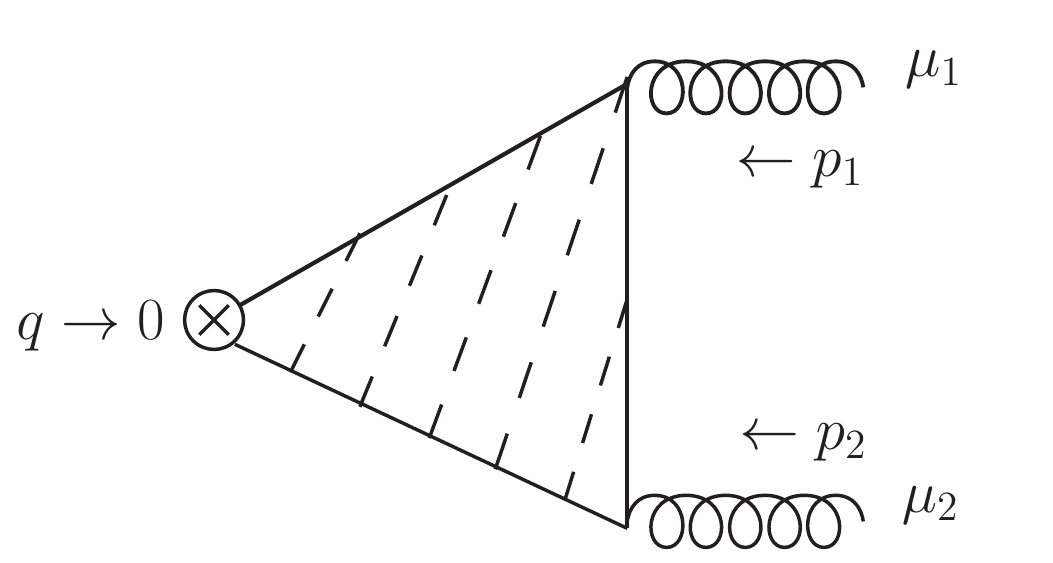}
\caption{The kinematic configuration chosen for the VVA diagrams.}
\label{fig:kinematics}
\end{center}
\end{figure}

In order to nullify possible IR divergences, as well as to utilize the analytic results for four-loop massless propagator-type master integrals~\cite{Baikov:2010hf,Lee:2011jt}, we choose to evaluate these matrix elements at a specific kinematic configuration~\cite{Bos:1992nd,Larin:1993tq}, illustrated in figure~\ref{fig:kinematics}.
The momentum insertion $q = -p_1-p_2$ through the axial-current vertex is zero, and the momenta $p_1=-p_2$ of the external gluons are taken off-shell. 
We project both $\Gamma^{\mu \mu_1 \mu_2}_{lhs}(p_1, -p_1)$ and $\Gamma^{\mu \mu_1 \mu_2}_{rhs}(p_1, -p_1)$ with the following projector devised in ref.~\cite{Ahmed:2021spj}, 
\begin{eqnarray}
\label{eq:anomalyprojector}
\mathcal{P}_{\mu \mu_1 \mu_2} = -\frac{1}{6 \, p_1 \cdot p_1} \, \epsilon_{\mu\mu_1\mu_2\nu}\, p_1^{\nu} \,.
\end{eqnarray}
This leads to the following scalar matrix elements: 
\begin{eqnarray}
\label{eq:smeL}
\mathcal{M}_{lhs} &=&  \mathcal{P}_{\mu \mu_1 \mu_2} \,  \Gamma^{\mu \mu_1 \mu_2}_{lhs}(p_1, -p_1) \,, \nonumber\\
\mathcal{M}_{rhs} &=&  \mathcal{P}_{\mu \mu_1 \mu_2} \,  \Gamma^{\mu \mu_1 \mu_2}_{rhs}(p_1, -p_1) \,.
\end{eqnarray}
In fact, eq.~\eqref{eq:anomalyprojector} encodes the one and only ``physical'' Lorentz structure of $\Gamma^{\mu \mu_1 \mu_2}_{lhs}(p_1, -p_1) $ surviving at the chosen kinematics $q=0,\, p_2 = - p_1$ under the condition of having one Levi-Civita tensor and being Bose-symmetric.
Consequently, the projector $\mathcal{P}_{\mu \mu_1 \mu_2}$ projects out the form factor in front of this unique structure, which is not vanishing in the limit $q=0$ and closely related to the anomaly of the axial current, albeit not very obvious from eq.~(\ref{eq:anomalyprojector}) at first sight. (We refer to refs.~\cite{Ahmed:2021spj,Chen:2022aqw} for details about this point.)
%There is no need to use the so-called physical polarization sum rule for these external gluons, because the momentum-dependent part makes no contribution after contracting with the structure in eq.~\eqref{eq:anomalyprojector} (namely this structure is transversal by itself).
The product of the two $\epsilon^{\mu\nu\rho\sigma}$ in eq.~\eqref{eq:smeL} is replaced by a combination of products of spacetime-metric tensors $g^{\mu\nu}$ according to eq.~\eqref{eq:LeviCivitaContRule}, and the dimensionality of the Lorentz indices of $g^{\mu\nu}$ is set to $D\neq 4$ as described at the end of section~\ref{sec:RoK}.
~\\

The work-fow and technical tools employed to evaluate the scalar matrix elements or form factors $\mathcal{M}_{lhs}$ and $\mathcal{M}_{rhs}$ defined above follow closely those used in previous calculations~\cite{Ahmed:2021spj,Chen:2021gxv}, and we refer in particular section 4 of ref.~\cite{Ahmed:2021spj} for details.
The only new addition in the computational set-up for the present study concerns the implementation of the $\g5$-trace in Kreimer scheme in \form, following exactly the algorithmic procedure formulated in section~\ref{sec:RoK}.

\subsection{The observation and discussion}
\label{sec:obs}

The leading $\mathcal{O}(a_s)$ result for the bare $\mathcal{M}_{lhs} = \sum_{n=1} \hat{a}_s^n\,  \mathcal{M}_{lhs}^{[n]}$ expanded in the bare coupling $\hat{a}$ reads, in our choice of normalization,
\begin{eqnarray}
\label{eq:Mlhs_1L}
\mathcal{M}_{lhs}^{[1]} =  n_f\, \mathrm{N}_{\epsilon}\, 
4\,\frac{e^{\epsilon \gamma_E} \,\Gamma(1+\epsilon) \, \Gamma^2(1-\epsilon) }{\Gamma(2-2\epsilon)}
\end{eqnarray}
where $\mathrm{N}_{\epsilon} \equiv 1 - \frac{11}{3} \epsilon + 4 \epsilon^2 - \frac{4}{3} \epsilon^3 $ is related to the squared norm of $\epsilon_{\mu\mu_1\mu_2\nu} p_1^{\nu}$ in eq.~\eqref{eq:anomalyprojector} with $p_1^2=1$, and $\gamma_E$ is the Euler constant.  
The color factor $\delta_{ab}$ (with $a\,, b$ the color indices of the two external gluons) is suppressed in eq.~\eqref{eq:Mlhs_1L}. 
$\mathcal{M}_{lhs}^{[1]}$ is finite and equals to $4\,n_f$ at $\epsilon = 0$. 
In the same normalization convention, the leading $\mathcal{O}(a_s^0)$ result for the bare $\mathcal{M}_{rhs} = \sum_{n=0} \hat{a}_s^n\,  \mathcal{M}_{rhs}^{[n]}$ reads
\begin{eqnarray}
\label{eq:Mrhs_0L}
\mathcal{M}_{rhs}^{[0]} = 8 \, \mathrm{N}_{\epsilon}\,,
\end{eqnarray}
which has the ABJ equation~\eqref{eq:ABJanomalyEQ} satisfied, albeit only in the limit $\epsilon = 0$.\footnote{We note that this fact will not be changed by using a Levi-Civita $\epsilon^{\mu\nu\rho\sigma}$ with strictly 4-dimensional Lorentz indices, and hence is not an artifact of the adopted prescription for $\epsilon^{\mu\nu\rho\sigma}$ described at the end of the section~\ref{sec:RoK}. 
Because the non-rational $\epsilon$ dependence in eq.~\eqref{eq:Mlhs_1L} arises from the dimensionally regularized loop integrals.}

The result for $\mathcal{M}_{lhs}^{[1]}$ is identical to the one obtained in ref.~\cite{Larin:1993tq,Ahmed:2021spj} with full $\epsilon$ dependence using Larin's variant of the non-anticommuting $\g5$~\cite{Larin:1991tj,Larin:1993tq}.
Because i) the $\oVFF$ in figure~\ref{fig:Max1PIopenVFF} for the one-loop VVA diagram reduces to the tree-level axial-coupling vertex, and consequently the axial-current matrix implicit in eq.~\eqref{eq:FCdef} becomes identical to eq.~\eqref{eq:gamma5-axial};  
ii) furthermore the Levi-Civita tensor $\epsilon^{\mu\nu\rho\sigma}$ is treated as described at the end of section~\ref{sec:RoK}.
Indeed, it is straightforward to see that with $\epsilon^{\mu\nu\rho\sigma}$ treated in this way, if one had always chosen the axial-current vertex as the reading-point in all VVA diagrams, e.g.~those in figure~\ref{fig:samplediagrams}, then the bare expression of $\mathcal{M}_{lhs}$ at each perturbative order would be completely identical to those~\cite{Larin:1993tq,Ahmed:2021spj} determined using Larin's prescription. 
In particular, the renormalization for the singlet axial-current operator determined under this reading prescription will also be the same as in Larin's prescription~\cite{Larin:1991tj,Larin:1993tq}, known to $\mathcal{O}(a_s^5)$ to date~\cite{Chen:2022lun}.
With $\g5$ treated in Kreimer scheme, this equality no longer holds for the full $\mathcal{M}_{lhs}$ beyond $\mathcal{O}(a_s)$. 
However, it still remains true at least for the class of diagrams with exactly the anomalous fermion-triangle and fermion-box subgraphs, such as the middle one in the first row and the right-most one in the second row of figure~\ref{fig:samplediagrams}, as for this kind of \textit{genuinely} axial-anomalous diagrams the $\oVFF$ is always the tree-level axial-coupling vertex. 
We have checked this point in our results for the bare $\mathcal{M}_{lhs}$. 
~\\

The bare $\mathcal{M}_{lhs}$ and $\mathcal{M}_{rhs}$ contain UV divergences, but are free of IR divergences due to the external gluons being taken off-shell. 
As mentioned above, our expressions for $\mathcal{M}_{lhs}$ determined in Kreimer scheme start to differ from that derived in Larin's convention from $\mathcal{O}(a_s^2)$, due to difference in the ``non-anomalous'' two-loop diagrams, e.g.~the right-most diagram in the first row of figure~\ref{fig:samplediagrams}. 
For $\mathcal{M}_{lhs}$ up to $\mathcal{O}(a_s^2)$, the ABJ anomalous triangle does not yet enter in the loop corrections to the quark axial-coupling vertex, which only appears at one order higher, e.g.~the rightmost diagram in the second row of figure~\ref{fig:samplediagrams} (which is known as the light-by-light correction to the axial anomaly in refs.~\cite{Anselm:1989gi,Larin:1993tq}). 
Consequently, according to ref.~\cite{Kreimer:1993bh}, the renormalization needed for $\mathcal{M}_{lhs}$ up to $\mathcal{O}(a_s^2)$ is expected to be very simple: the renormalization of the coupling $\alpha_s$ and of the external gluon fields in $\MSbar$ scheme determined in QCD with $n_f$ massless quarks using the $\xi$-gauge should be sufficient. 

To be more explicit, according to eq.\eqref{eq:ABJNaiveBare} and eq.\eqref{eq:ABJNaiveRen}  in Kreimer scheme, $\mathcal{M}_{lhs}$ and $\mathcal{M}_{rhs}$ are \textit{expected} to be renormalized as follows: 
\begin{eqnarray}
\label{eq:KMuvr}
&&\mathrm{M}_{lhs} \equiv Z^{A_5}_s\, Z_3 \, \mathcal{M}_{lhs}\left( {\hat a}_s\,,\, \hat{\xi} \right) =  Z^{A_5}_s\, Z_3 \, \mathcal{M}_{lhs}\left( Z_{a_s}\, a_s\,,\, 1 - Z_3 + Z_3\, \xi  \right) = \sum_{n=1}^{\infty} a_s^n\,  \mathrm{M}_{lhs}^{[n]} \,, \nonumber\\
&& \mathrm{M}_{rhs} \equiv
Z_{F\tilde{F}} \, Z_3 \, \mathcal{M}_{rhs}\left( {\hat a}_s\,,\, \hat{\xi} \right) 
\,+\, Z_{KJ} \, Z_3 \, \mathcal{M}_{lhs}\left( {\hat a}_s\,,\, \hat{\xi} \right) = \sum_{n=1}^{\infty} a_s^n\,  \mathrm{M}_{rhs}^{[n]} \,, 
\end{eqnarray}
where we have suppressed the kinematic dependence in $\mathcal{M}_{lhs/rhs}$ as well as their renormalized counterparts $\mathcal{M}_{lhs/rhs}$.
Two additional quick comments are in order regarding the renormalization formula~\eqref{eq:KMuvr} used in this section.
First, since the external gluons are set off-shell, we need to include the corresponding $\MSbar$ wavefunction renormalization constant $Z_3$. 
Second, $\mathcal{M}_{lhs/rhs}$ projected out using eq.~\eqref{eq:anomalyprojector} are evaluated at an off-shell kinematics, illustrated in figure~\ref{fig:kinematics}, depend on the covariant-gauge fixing parameter $\xi$, which by itself requires renormalization in QCD. 
With $\xi$ defined through the gluon propagator $\frac{i}{k^2} \left(-g^{\mu\nu} + \xi \,\frac{k^{\mu} k^{\mu}}{k^2} \right)$, its renormalization reads $\hat{\xi} = 1 - Z_3 + Z_3\, \xi $ where one has used the fact that $1-\hat{\xi}$ renormalizes multiplicatively with the renormalization constant $Z_{\xi}=Z_3$ (see, e.g.~\cite{Czakon:2004bu,Chetyrkin:2004mf}).
(We refer to refs.~\cite{Larin:1993tq,Ahmed:2021spj} for a very detailed exposition of the standard part of the renormalizations for $\mathcal{M}_{lhs}$ and $\mathcal{M}_{rhs}$)

The so-renormalized $\mathrm{M}_{lhs} = \sum_{n=1} a_s^n\,  \mathrm{M}_{lhs}^{[n]}$ is indeed finite at $\mathcal{O}(a_s^2)$, which reads 
\begin{eqnarray}
\label{eq:mlhs_2L}
\mathrm{M}_{lhs}^{[2]} =  n_f\, \big(C_A\, (24 \,+\, 8\, \xi\, -\, 2\, \xi^2 \big) \big) \,+\, 4\, n_f\, C_F\,.
\end{eqnarray}
The definition of the quadratic Casimir color constants involved in this result is as usual: $C_A = N_c \,, \, C_F = (N_c^2 - 1)/(2 N_c) \,$ with $N_c =3$ in QCD and the color-trace normalization factor $\TR = {1}/{2}$.
%The general-covariant-gauge fixing parameter $\xi$ is defined through the gluon propagator $\frac{i}{k^2} \left(-g^{\mu\nu} + \xi \,\frac{k^{\mu} k^{\mu}}{k^2} \right)$, $k$ being the momentum of the gluon. 
%With this parametrization, the Feynman gauge corresponds to $\xi = 0$. 

However, to our surprise, $\mathrm{M}_{lhs}^{[2]}$ in eq.~\eqref{eq:mlhs_2L} does not equal to the $\mathcal{O}(a_s^2)$ result of $a_s\, n_f\, \TR \,\mathrm{M}_{rhs}$, the corresponding matrix element of the r.h.s.~of the ABJ equation~\eqref{eq:ABJanomalyEQ}, which reads 
\begin{eqnarray}
\label{eq:mrhs_1L}
n_f \, \TR\, \mathrm{M}_{rhs}^{[1]} =  n_f\, \big(C_A\, (24 \,+\, 8\, \xi\, -\, 2\, \xi^2 \big) \big) \,,
\end{eqnarray}
%where $\TR = 1/2$ is inserted on the r.h.s., 
differing by a finite amount $4\, n_f\, C_F$. 
More explicitly, the discrepancy between the two renormalized matrix elements defined in eq.~\eqref{eq:KMuvr} up to $\mathcal{O}(a_s^2)$ reads 
\begin{eqnarray}
\label{eq:diff_1L}
\mathrm{M}_{lhs} - a_s\, n_f \, \TR\, \mathrm{M}_{rhs} \Big|_{\epsilon =0} 
= a_s^2\,\big( 4\, n_f\, C_F \big) \,+\, \mathcal{O}(a_s^3)\,. 
\end{eqnarray}
Note that $\mathrm{M}_{rhs}^{[1]}$ in eq.\eqref{eq:mrhs_1L} is completely independent of the $\g5$-prescription, 
which involves only gluon self-interactions (e.g.~the middle diagram in the second row of figure~\ref{fig:samplediagrams}), and thus can be determined unambiguously free of any issue related to different $\g5$-prescriptions.
In QED, $\mathrm{M}_{lhs}^{[2]}$ reduces to $2\,n_f$ which is not zero, 
while $\mathrm{M}_{rhs}^{[1]}$ equals to 0!
In this simpler case, it is easier to locate the source of this discrepancy to be precisely the treatment of the right-most two-loop diagram in the first row of figure~\ref{fig:samplediagrams} in Kreimer scheme.
~\\

By inspecting the results at hand, we find that this discrepancy can only be amended by introducing an extra additive part, proportional to the Chern-Simons current $K^{\mu}$, in the renormalization of an external singlet axial-current operator $J^{\mu}_{A_5,s} \equiv \sum_{i=1}^{n_f} \, \bar{\psi}^{B}_{i} \,\gamma^{\mu} \g5 \, \psi^{B}_i$ in QCD defined with $\g5$ treated in Kreimer scheme, 
\begin{eqnarray}
\label{eq:J5ACR}
\big[J^{\mu}_{A_5,s}\big]_R = Z^{A_5}_s\, \big[J^{\mu}_{A_5,s}\big]_B \,+\, 
Z_{JK} \, K^{\mu} 
\end{eqnarray}
on top of the purely multiplicative part $Z^{A_5}_s\, \big[J^{\mu}_{A_5,s}\big]_B$.\footnote{There is actually another, albeit minor, issue related to the proper definition of this purely multiplicative part, which however becomes relevant only from $\mathcal{O}(a_s^3)$ to be discussed later in this section. %This may be compared with the renormalization of the $K^{\mu}$ defined by eq.~\eqref{eq:Kcurrent}.
}
Perturbative QCD corrections to $Z^{A_5}_s$ start from $\mathcal{O}(a_s^2)$, solely due to the anomalous subgraphs, and are not yet relevant for the discussion of the vacuum-gluon matrix elements above.
Consequently, the above expressions allow us to determine the following result: 
\begin{eqnarray}
\label{eq:zjk_LO}
Z_{JK} = -\frac{1}{2} n_f \, C_F \, a_s^2 \,+\, \mathcal{O}(a_s^3) \,.
\end{eqnarray} 
If one now updates the definition of $\mathrm{M}_{lhs}$ in eq.\eqref{eq:Kcurrent} by using the singlet axial-current operator renormalized as in eq.\eqref{eq:J5ACR} with the result for $Z_{JK}$ given in eq.\eqref{eq:zjk_LO}, then one has  
$\mathrm{M}_{lhs} - a_s\, n_f \, \TR\, \mathrm{M}_{rhs} \Big|_{\epsilon =0} = \mathcal{O}(a_s^3)\,$ as expected.

Apart from defying the multiplicative renormalization of the axial-current operator as in eq.~\eqref{eq:Zsmatrix}, eq.~\eqref{eq:J5ACR} is in tension, at least formally, with the gauge-invariance of a singlet axial-current operator.
According to the general theory on renormalization of gauge-invariant operators~\cite{Dixon:1974ss,Joglekar:1975nu,Kluberg-Stern:1974iel,Kluberg-Stern:1974nmx,Kluberg-Stern:1975ebk,Joglekar:1976eb,Collins:1984xc,Joglekar:1976pe,Henneaux:1993jn,Barnich:1994ve} the so-called unphysical operators that may get involved in the renormalization of a (physical) gauge-invariant operator shall be either BRST-exact, i.e.~BRST variation of certain \textit{local} operators, or vanishing by the equation-of-motion, both having vanishing matrix elements between on-shell physical states. 
The Chern-Simons current $K^{\mu}$ itself is gauge-variant but not a BRST-exact operator, i.e.~not BRS variation of any \textit{local} operator (see, e.g.~\cite{Bertlmann:1996xk,Bilal:2008qx}), and does not belong to the class of unphysical operators mentioned above.
We thus attribute the unusual behavior in eq.~\eqref{eq:J5ACR} to a mere technical consequence of the special $\g5$ manipulation in Kreimer scheme when applied to the anomalous diagrams.  
Recall that the ABJ anomalous fermion loop in dimensional regularization is special regarding the following aspect: the degree of its superficial UV divergence is zero rather than negative, while the absence of an overall UV divergence is a consequence of subtle cancellation of intermediate spurious poles between different trace terms which are sensitive to the $\g5$ position in the fermion chain. 
On the other hand, from the pure diagrammatic point of view, %especially the BPHZ theorem (putting aside the gauge-invariance argument which is not really there due to the gauge-fixing term, which is weakened into the so-called BRST-symmetry/invariance), 
the absence of an overall UV divergence in the VVA diagrams %despite its non-negative degree of superficial UV divergence 
at any perturbative order is crucial for the multiplicative renormalization of the singlet axial-current operator, i.e.~the absence of the mixing with the K-current operator under renormalization.
The aforementioned discrepancy between eq.~\eqref{eq:mlhs_2L} and eq.~\eqref{eq:mrhs_1L}, spelled out explicitly in eq.\eqref{eq:diff_1L}, seems to imply that shifting $\g5$ anticommutatively from the axial vertex to I/O-legs according to Kreimer prescription~\cite{Kreimer:1993bh} also introduces (potentially divergent and gauge-dependent) spurious anomalous terms, albeit only for VVA-type diagrams. 
Indeed we see no such kind of issue in the matrix elements of a pseudo-scalar operator between the vacuum and a pair of external gluons to be presented in the next section.
By ``VVA-type'' used here and in the text below, we refer to a subgraph with one external (color-neutral) axial-current operator and two or three gluon legs, essentially the configurations where the Chern-Simons current operator~\eqref{eq:Kcurrent} can have non-vanishing tree-level matrix elements.
~\\

One direct implication of the above observation is that the calculation of axial-anomalous diagrams using Kreimer scheme is more subtle than anticipated in the original ref.~\cite{Kreimer:1993bh}.
In particular, our finding eq.~\eqref{eq:J5ACR} means that this scheme itself does not offer a constructive proof for the Adler-Bell-Jackiw anomaly equation or the related Adler-Bardeen theorem. 
Instead, demanding the validity of this equation, as an external physical input, would force one to manually reintroduce certain additional compensation terms in this scheme, just like in schemes based on a non-anticommuting $\g5$, which were intended to be avoided by the very invention of this scheme!
~\\

However, this is not yet the end of the story. 
For $\mathrm{M}_{lhs}$ starting from $\mathcal{O}(\alpha^3_s)$, another potential issue with using Kreimer scheme begins to emerge as well, which is in a sense less un-expected compared to the one reported above.
This issue can be best indicated by the following question:
how to determine the mixing coefficient in front of the axial-current operator in the additive renormalization of the Chern-Simons current $K^{\mu}$, c.f.~eq.~\eqref{eq:Zsmatrix}, and subsequently ensure its universality, with $\g5$ treated according to Kreimer scheme?\footnote{As for as the computations presented in this work are concerned, only the leading term $a_s\, 12\, C_F/\epsilon$ for this mixing coefficient $Z_{KJ}$ in eq.\eqref{eq:KMuvr} is needed which reads the same as in the Larin's scheme~\cite{Larin:1993tq}.
Difference between the two may appear at high orders, which needs to be checked explicitly.}  
The point behind this question is that the $\g5$ or axial-current operator in these UV counterm-terms are \textit{secondary}: they are induced by combining the Levi-Civita tensor $\epsilon^{\mu\nu\rho\sigma}$ with a fermion chain of a UV divergent subgraph that is primarily free of any axial vertex, rather than inserted from the outset by the Feynman rules of the theory.
However, there can be ambiguity in the defining expression for a fermion chain in Kreimer scheme depending on whether this kind of \textit{would-be} $\g5$ is explicitly manifested or not in the treatment.\footnote{One may compare this with the remarks in ref.~\cite{Chetyrkin:1997gb} and in the manual of \form~\cite{Vermaseren:2000nd} on the usage of the so-called reduction formula, which rewrites a triple product of elementary Dirac matrices in terms of $\g5$ by virtue of the 4-dimensional Chisholm identity, in traces not meant to be done in 4 dimensions. %Treating the $\g5$ so-induced by Kreimer scheme in general would only make the matter more messy.
}

In view of this origin, on one hand, a secondary $\g5$ induced in UV counter-terms would be naturally treated as ``non-anticommuting'' in order to achieve a consistency in the treatment of the relevant fermion chains between these UV counter-terms and the original bare diagrams (where the corresponding fermion chains are detached from the $K$-current or anomalous axial-current vertex).
On the other hand, in the same spirit of Kreimer scheme, we would like to eliminate the spurious anomalous pieces generated by embedding $\g5$-vertex within loop corrections as much as possible. 
Namely, we do not want to have in the matrix elements of a $\g5$-dependent operator any spurious anomalous pieces due to the apparent loss of $\g5$'s anticommutativity, otherwise one could have rolled back to a non-anticommuting-$\g5$ prescription completely from the outset.  
Based on the discussion presented in section~\ref{sec:RoK}, we thus propose the following compromise, an ad hoc \textit{invisible}-$\g5$ rule. 
\begin{itemize}
\item 
To distinguish, we first recycle the notation $\hat{\g5}$ introduced in eq.~\eqref{eq:gamma5} and use it to denote the \textit{invisible}-$\g5$ in the axial-current operator induced in UV counter-terms. 
When computing these UV counter-terms, right after shifted anticommutatively to the I/O-leg of $\oVFF$, this $\hat{\g5}$ will be immediately dissolved from the fermion chain by substituting the l.h.s.~of eq.~\eqref{eq:gamma5} in terms of the Levi-Civita tensor, hence becoming ``invisible'' and no more further operations regardless of whether there is any normal $\g5$ inserted by the Feynman rules in the remaining part of this fermion chain.
%Afterwords, it will neither be moved further away from these positions nor these positions be taken as reading-points for the explicit $\g5$ inserted by the Feynman rules, regardless of the diagrams which it is embedded in or interfered to.
\end{itemize} 

By virtue of the aforementioned shifting, the only possible source of the UV counter-terms needed for an external singlet axial-current operator is the Feynman diagrams with VVA-type subgraphs, namely a $\g5$-odd fermion triangle or box loop with the singlet axial-current operator and possible virtual corrections as defined above. 
The crucial point in the above ad hoc rule is that this $\hat{\g5}$ or \textit{invisible}-$\g5$ will not be manifested explicitly like a normal $\g5$ in the fermion chain, in order to ensure a consistency in the treatment of this fermion chain between the UV counter-terms and the corresponding bare diagrams. 
Once again, this reintroduces a non-anticommuting $\g5$ to some extent in this scheme.
Taking this into account, the renormalization formula~\eqref{eq:J5ACR} for an external singlet axial-current operator in this Kreimer-scheme variant may be further revised as 
\begin{eqnarray}
\label{eq:J5ACRR}
\Big[
\sum_{i=1}^{n_f} \, \bar{\psi}^{B}_{i}  \,\gamma^{\mu} \g5 \, \psi^{B}_i
\Big]_R 
= Z^{A_5}_s\,  \sum_{i=1}^{n_f} \, \bar{\psi}^{B}_{i}  \,\gamma^{\mu} \hat{\g5} \, \psi^{B}_i \,+\, Z_{JK} \, K^{\mu} 
\end{eqnarray}
in order to be applicable unambiguously beyond $\mathcal{O}(a_s^2)$ in QCD. 
%Note that the $\hat{\g5}$ in the r.h.s.~of eq.\eqref{eq:J5ACRR} will be treated according to the above an ad hoc \textit{invisible}-$\g5$ rule, rather than like in a non-anticommuting $\g5$ scheme, and hence there is no problem with the Furry's theorem.
Although this add hoc rule may not be very elegant in practice, at least it offers an unambiguous framework to demonstrate the key message offered by this investigation: 
the renormalization of the singlet axial-current operator with $\g5$ treated in Kreimer scheme is not strictly multiplicative, but takes a more involved form~\eqref{eq:J5ACRR}, and the ABJ equation does not hold automatically in the bare form~\eqref{eq:ABJNaiveBare} in this kind of schemes.
In other words, different from the claim in the original ref.~\cite{Kreimer:1993bh}, we demonstrate that the Kreimer prescription itself does not offer a constructive proof for the Adler-Bardeen theorem~\cite{Adler:1969gk}.

Furthermore, one may expect to encounter similar kind of issues, if not more involved, when trying to apply the original Kreimer scheme, as well as its variant discussed above, to an effective field theory where the axial-current or pseudo-scalar operators may appear in the effective Lagrangian or mixed renormalizations of local-composite operators, which are of course beyond the scope envisaged initially in the original ref.~\cite{Kreimer:1993bh}.
~\\

By demanding the validity of the ABJ equation at least for renormalized operators, which translates into the equality between the revised $\mathrm{M}_{lhs}$ defined using eq.\eqref{eq:J5ACRR} and $ a_s\, n_f\, \TR \,\mathrm{M}_{rhs}$ in the limit $\epsilon = 0$ (assuming absence or proper subtraction of IR divergences), the following result for $Z_{JK}$ up to $\mathcal{O}(a_s^3)$ can be determined: 
{\small 
\begin{eqnarray}
\label{eq:zjk_NLO}
Z_{JK} &=& a_s^2\, \Big( -\frac{1}{2} n_f \, C_F \Big)  
 \,+\, 
a_s^3 \, \Big\{ \,
\frac{1}{\epsilon}\, \Big( -\frac{4}{9}\,n_f^2\,C_F \,+\, n_f \, C_A\, C_F\, \big(\frac{97}{36} - \frac{\xi}{24}\big) \Big) \nonumber\\
&+& 
\Big( 
\frac{5}{54}\, n_f^2\,C_F  +
 n_f \Big( C_F^2 \big(\frac{4}{3} + \frac{\xi}{3} - \frac{\xi^2}{6} \big) + 
C_A C_F \big(\frac{109}{216} - \frac{53 \xi}{48} + \frac{\xi^2}{6} \big) \Big)
\Big) \Big\} +
\mathcal{O}(a_s^4) .
\end{eqnarray}
}
Note that the $a_s^3$ coefficient of $Z_{JK}$ starts to develop poles in $\epsilon$ as well dependence on the gauge-fixing parameter $\xi$.
To arrive at this result, one needs the input $Z^{A_5}_s = 1 \,+\, a_s^2\, n_f \, C_F \big( \frac{3}{\epsilon}  + \frac{3}{2} \big)\,+\, \mathcal{O}(a_s^3)$, and $Z_{KJ} = a_s\, \frac{12\, C_F}{\epsilon} \,+\, \mathcal{O}(a_s^2)$.
This $\mathcal{O}(a_s^2)$ result for $Z^{A_5}_s$ can be determined unambiguously by demanding the validity of the ABJ equation when inserted between the vacuum and the state of a pair of external quarks up to two-loop order, where the $Z_{JK}$ is not yet involved. 
Up to $\mathcal{O}(a_s^2)$, the $Z^{A_5}_s$ agrees with the so-called ``pure singlet'' renormalization constant~\cite{Chetyrkin:1993hk,Chetyrkin:1993jm,Chetyrkin:1993ug,Bernreuther:2005gw,Gehrmann:2021ahy,Chen:2021rft}.
At this point, it is worthy to mention that $Z^{A_5}_s - a_s\, n_f\, \TR \, Z_{KJ} \neq 1$ is observed in Kreimer scheme, which is explicitly verified up to $\mathcal{O}(a_s^2)$ in this calculation.
In Kreimer scheme, the perturbative corrections to both $Z^{A_5}_s$ and $Z_{JK}$ start from $\mathcal{O}(a_s^2)$, all due to the presence of the VVA-type subgraphs in Feynman diagrams. 
From the above discussion, we see that the VVA-type diagrams in dimensional regularization once more demonstrate their peculiarity in defying a simple and elegant treatment which would seem to be very natural to us. 

\subsection{Checks on the AWI with massive quarks}
\label{sec:Calc-onshell}

In this section we discuss briefly a few checks related to the observation presented in the preceding section in order to clarify a few obvious questions which one may naturally have in mind. For example, one may be skeptical about the special kinematics and also the projector in use, wondering whether they could be responsible for the strange behavior, whether there is this kind of issue if one replaces the axial-current operator by a pseudo-scalar operator and also whether for non-anomalous amplitudes Kreimer scheme works again without any additional manual tweaks.~\\

To get the answers to these questions, we computed the matrix elements of both the (current) divergence of the singlet axial current $\partial_{\mu} J^{\mu}_{A_5,s}$ and the pseudo-scalar operator $J_{ps} \equiv \sum_{j=1}^{n_f} m_j \, \bar{\psi}_{j}  \, i\gamma_5 \, \psi_j$ between the vacuum and a pair of on-shell gluons in QCD with $n_f=2$ (massive) quark flavors at on-shell kinematics: 
\begin{eqnarray}
\label{eq:kineOS}
p_1^2 = p^2_2 = 0\,, \quad (p_1+p_2)^2 = (-q)^2 = s\,.
\end{eqnarray}
Keeping the quark propagators massive is necessary both for non-vanishing vacuum-gluon matrix elements for $J_{ps}$ and also to obtain non-vanishing results for a non-anomalous or non-singlet axial-current combination.
The definitions of the aforementioned matrix elements are in close analogy with those in section~\ref{sec:Calc-offshell}, c.f.~eqs.~\eqref{eq:Glhs1PI},~\eqref{eq:Grhs1PI}, 
specifically,  
\begin{eqnarray}
\label{eq:Gano1PI}
\Gamma^{\mu_1 \mu_2}_{A_5,s}(p_1, p_2, m_Q) &\equiv&  
\int d^4 x  d^4 y \, e^{-i p_1 \cdot x - i q \cdot y }\,  \langle 0| \hat{\mathrm{T}}\left[\partial_{\mu} J^{\mu}_{A_5,s}(y) \, A_a^{\mu_1}(x) \, A_a^{\mu_2}(0) \right] |0 \rangle |_{\mathrm{amp}}\,, \nonumber\\
\Gamma^{\mu_1 \mu_2}_{ps}(p_1, p_2, m_Q) &\equiv&  
\int d^4 x  d^4 y \, e^{-i p_1 \cdot x - i q \cdot y }\,  \langle 0| \hat{\mathrm{T}}\left[J_{ps}(y) \, A_a^{\mu_1}(x) \, A_a^{\mu_2}(0) \right] |0 \rangle |_{\mathrm{amp}}\,,  
\end{eqnarray}
where $m_Q$ denotes the non-zero mass of the internal quark propagator.
With these ingredients at hand, we can check the so-called Axial Ward Identity (AWI) following from inserting 
\begin{eqnarray} 
\label{eq:ABJanomalyEQ_massive}
\big[\partial_{\mu} J^{\mu}_{A_5,s} \big]_{R} &=& a_s\, n_f\, \TR \,  \big[F \tilde{F} \big]_{R}\,
+\, 2\, \big[J_{ps}\big]_{R} \nonumber\\
\big[\partial_{\mu} J^{\mu}_{A_5,ns} \big]_{R} &=& 2\, \Big[\sum_{j=1}^{n_f} \lambda_j\, m_j \, \bar{\psi}_{j}  \, i\gamma_5 \, \psi_j \Big]_{R} 
\end{eqnarray}
between on-shell matrix elements (see, e.g.~\cite{Adler:1969gk,Itzykson:1980rh,Bernreuther:2005rw}).
The second one in eq.~\eqref{eq:ABJanomalyEQ_massive} is the AWI for a non-singlet axial-current operator $J^{\mu}_{A_5,ns}$ where $\lambda_j$ on the r.h.s.~denotes the (flavor) isospin of the quark field $\psi_j$.

At the standard kinematics~\eqref{eq:kineOS}, it is well-known that there is just one Lorentz tensor structure%  
\begin{eqnarray}
\label{eq:anomalyprojectorOS}
\mathcal{P}_{\mu_1 \mu_2} = \epsilon_{\mu_1\,\mu_2 \,p_1\, p_2}\,
\end{eqnarray}
both in $\Gamma^{\mu_1 \mu_2}_{A_5,s}(p_1, p_2, m_Q)$ and $\Gamma^{\mu_1 \mu_2}_{ps}(p_1, p_2, m_Q)$ under the condition of having one Levi-Civita tensor and being Bose-symmetric. % which are respected in Kreimer scheme. 
Like in eq.~\eqref{eq:smeL}, we extract the respective (un-normalized) form factors of $\Gamma^{\mu_1 \mu_2}_{A_5,s}(p_1, p_2, m_Q)$ and $\Gamma^{\mu_1 \mu_2}_{ps}(p_1, p_2, m_Q)$ in front of this unique transversal Lorentz tensor structure simply by contracting them with eq.~\eqref{eq:anomalyprojectorOS}, namely, 
\begin{eqnarray}
\label{eq:anomalyOS}
\Gamma_{A_5,s}(s, m_Q) &\equiv& \mathcal{P}_{\mu_1 \mu_2}\, \Gamma^{\mu_1 \mu_2}_{A_5,s}(p_1, p_2, m_Q), \nonumber\\
\Gamma_{ps}(s, m_Q) &\equiv& \mathcal{P}_{\mu_1 \mu_2}\,  \Gamma^{\mu_1 \mu_2}_{ps}(p_1, p_2, m_Q)\,.
\end{eqnarray}
%There is no need to use the so-called physical polarization sum rule for these external gluon, because the momentum-dependent part makes no contribution after contracting with the structure in eq.~\eqref{eq:anomalyprojectorOS} (namely this structrue is transversal by itself).

The same computational set-up and work-flow are employed here as in the calculation discussed in the proceeding sections. 
The main difference concerns the evaluation of the two-loop massive master integrals in $\Gamma_{A_5,s}(s, m_Q)$ and $\Gamma_{ps}(s, m_Q)$, which we choose to evaluate numerically using the AMF-method~\cite{Liu:2017jxz,Liu:2020kpc,Liu:2021wks,Liu:2022mfb,Liu:2022chg}. 
Compared to the previous massless calculation, one should be careful with the generation of the mass counter-terms in Kreimer scheme.
This is because, in the language of the non-cyclic $\g5$-trace~\cite{Kreimer:1989ke,Korner:1991sx,Kreimer:1993bh} different reading points could lead to different trace expressions in D dimensions, or equivalently in the formulation given in section~\ref{sec:RoK} the constructively-defined $\hat{\g5}$ does not anticommute with the quark propagators.
After one has identified the I/O-leg of the $\oVFF$ subgraph, subsequently tagged by inserting a place-holder at either the head of I-leg or the tail of O-leg, one can then apply the usual Taylor expansion to each massive quark propagator. 
The guiding principle is simply that the mass counter-terms generated on the I/O-leg fermion propagators shall be placed \textit{outside} the $\oVFF$ subgraph, in order to be consistent with the 1PI condition of the latter.
%The mass counter-terms for the fermion propagators generated in the above way is still equivalent to the one derived through taking derivatives of the \textit{pre-calculated} bare amplitudes w.r.t the bare mass.
~\\

Suppressing the computational details, we now enter directly the discussion of the outcome of a few checks based on the calculation of the above quantities.  
\begin{itemize}
\item 
Keeping just one massive quark flavor for simplicity and clarity, we find that the IR-subtracted finite remainder of the on-shell renormalized form factor $\Gamma_{A_5,s}(s, m_Q)$ at $\mathcal{O}(a_s^2)$ determined in Kreimer scheme does not agree with the well established result determined using a non-anticommuting $\g5$ scheme. 
%To be specific, we take the Larin's variant/convention for the later. Actually the only $\g5$-related counter-term involves only the $\mathcal{O}(a_s)$ term of $Z_s$ which is the same between HV and Larin.
However, the discrepancy can again be accounted for by the additional renormalization term derived using eqs.~\eqref{eq:J5ACR},~\eqref{eq:zjk_LO}. 
This shows that this revised renormalization formula for the singlet axial-current operator defined in Kreimer scheme holds independent of the quark masses at least up to $\mathcal{O}(a_s^2)$. 

\item 
In the case of the pseudo-scalar form factor $\Gamma_{ps}(s, m_Q)$ at $\mathcal{O}(a_s^2)$, we find a perfect agreement in the results for its IR-subtracted finite remainder between Kreimer scheme and a non-anticommuting $\g5$ scheme.
In particular, in the calculation of $\Gamma_{ps}(s, m_Q)$ in Kreimer scheme we do not need to use any additional $\g5$-related renormalization constants, such as those given in ref.~\cite{Larin:1993tq}, other than those for $a_s$, external gluon fields and quark masses in QCD all needed just to $\mathcal{O}(a_s)$.
Therefore, there is no similar issue to that in section~\ref{sec:obs} for $\Gamma_{ps}(s, m_Q)$, which however does appear in $\Gamma_{A_5,s}(s, m_Q)$.

Furthermore, with the on-shell renormalization for the quark masses in the pseudo-scalar operator $J_{ps} \equiv \sum_{j=1}^{n_f} m_j \, \bar{\psi}_{j}  \, i\gamma_5 \, \psi_j$, we confirm that only after incorporating the additional renormalization term determined using eqs.~\eqref{eq:J5ACR},~\eqref{eq:zjk_LO} for $\Gamma_{A_5,s}(s, m_Q)$, will then the singlet AWI, the first one in eq.~\eqref{eq:ABJanomalyEQ_massive}, be preserved in Kreimer scheme.  

\item 

Once one considers the non-anomalous combination of two $\Gamma_{A_5,s}(s, m_Q)$ with different quark masses, which amounts to the vacuum-gluon form factor corresponding to a non-anomalous non-singlet axial-current operator, one observes that there is no more need of any additional $\g5$-related renormalization constant in order to preserve the non-singlet AWI, the second one in eq.~\eqref{eq:ABJanomalyEQ_massive}, in Kreimer scheme, which is checked explicitly to $\mathcal{O}(a_s^2)$.
%In other words, the spurious terms generated by shifting anticommutatively the $\g5$ from the original axial-current vertex to the I/O-legs within a VVA-type subgraph cancel in the total result for the non-anomalous quantities. 
This is expected, because the structure of these extra terms generated by shifting anticommutatively $\g5$ from the original axial-current or pseudo-scalar vertex to the I/O-legs within a VVA-type subgraph is reflected in the additional renormalization formula~\eqref{eq:J5ACRR}, closely related to the axial anomaly which cancels in the non-anomalous combination of form factors or amplitudes. 
A more systematic demonstration of this point can be made along the line of the discussion of the renormalization of the singlet and non-singlet type contributions to the quark form factors in refs.~\cite{Chetyrkin:1993hk,Chen:2021rft}, c.f.~eqs.(4.3),~(4,7) of ref.~\cite{Chen:2021rft}.

\end{itemize}

We conclude this section by a quick comment related to a remark in ref.~\cite{Bednyakov:2015ooa} on the operation of replacing one $\gamma^{\mu} \g5$ %, not its summarized form,
in a $\g5$-odd trace by the r.h.s.~of eq.~\eqref{eq:gamma5-axial}, referred to as ``Larin-like prescription'' therein. 
The remark was made in the context of the reading prescription of ref.~\cite{Korner:1991sx}, and indeed the charge-conjugation property of the diagrams considered in ref.~\cite{Bednyakov:2015ooa} may not be maintained properly in case $\gamma^{\mu}$ in this replacement comes from a fermion propagator rather than a gauge-interaction vertex.
From our point of view, the reason why this replacement shall be avoided in general in Kreimer scheme~\cite{Kreimer:1993bh} is that it may effectively entail a contribution where $\g5$ is placed inside, rather than outside, the non-singlet vertex correction $\oVFF$ (illustrated in figure~\ref{fig:Max1PIopenVFF}) if $\gamma^{\mu}$ in this replacement happens to come from a gauge-interaction vertex and $\oVFF$ is not a tree-level vertex.
Consequently this may introduce spurious anomalous terms which call for additional compensation terms just like in a non-anticommuting $\g5$ scheme. 
In case this $\gamma^{\mu}$ comes just from the momentum part in the numerator of the I/O-leg propagator, this replacement might be fine for some simple and normal diagrams, e.g.~those free of the VVA-type subgraphs, at least in view of the fact that they neither appear inside the subgraph $\oVFF$ nor violate Furry's theorem if applied to \textit{both} the I- and O-leg. %%For the VVA-type subgraphs one can easily check that the vector-current conservation on the two external vector-coupling vertices are not respected by the effective reading prescription resulting from this replacement; More explicit checks show that even for the non-anomalous combinations there seems to be something incorrect in the non-Abelian part.

\section{Conclusion}
\label{sec:Conc}

Calculating correctly the high-order perturbative corrections to quantities involving the singlet axial-current operator in dimension regularization is non-trivial due to the well-known $\g5$ issue in D dimensions. 
Kreimer scheme (according to the formulation in ref.~\cite{Kreimer:1993bh}) is a very neat and promising $\g5$ prescription in dimensional regularization, especially concerning its potential application to perturbative corrections in the full SM which may otherwise look quite daunting for other $\g5$ schemes.
Motivated by the question whether the Adler-Bell-Jackiw anomaly equation holds automatically in its bare form with an anticommuting $\gamma_5$ defined in Kreimer scheme, we calculated the matrix elements of an external singlet axial-current operator between the vacuum and a pair of gluons in QCD up to $\mathcal{O}(\alpha_s^3)$, which are known to be not vanishing. 
To this end, we have reformulated the original Kreimer scheme, and our variant, in terms of the standard Dirac trace with a constructively-defined non-anticommuting $\g5$ to facilitate an easy implementation in our computational set-up, where the treatment of the Levi-Civita tensor is slightly different from the original prescription (A general discussion is provided in the Appendix). 
We expect that this reformulation can be helpful to some practitioners who would like to try out the anticommuting $\gamma_5$ scheme using the public efficient tools for Dirac algebra.
Moreover, we propose a novel extension of this procedure, with the detail given in the Appendix, for further applications to the non-anomalous amplitudes in a renormalizable anomaly-free chiral gauge theory, e.g.~the electroweak theory, where in particular the symbolic manipulation of Levi-Civita tensor in DR is expected to be both unambiguous and technically advantageous.

To our surprise, we find that additional renormalization counter-terms proportional to the Chern-Simons current operator are needed for an external singlet axial-current operator starting from $\mathcal{O}(\alpha_s^2)$ in the variant of Kreimer scheme in use, and the same statement shall apply to the original Kreimer scheme too. 
This is in contrast to the well-known purely multiplicative renormalization for this operator defined with a non-anticommuting $\g5$. 
Consequently, without introducing compensation terms in the form of additional renormalization, the Adler-Bell-Jackiw anomaly equation does not hold automatically in the bare form in Kreimer scheme. 
This implies that this scheme itself does not offer a constructive proof for the related Adler-Bardeen theorem. 
We show, furthermore, that demanding the validity of this equation as an external physical input would force one to reintroduce some additional ad hoc rule in the calculation of the axial-anomalous Feynman diagrams in this scheme at high orders in perturbation.
We determine the corresponding gauge-dependent coefficient to $\mathcal{O}(\alpha_s^3)$ in QCD, using the aforementioned variant of the original Kreimer prescription which is implemented in our computation in terms of the standard cyclic trace together with a constructively-defined $\gamma_5$ and Lorentz indices of the Levi-Civita tensor taken D-dimensional.

In order to clarify a few obvious questions related to the aforementioned observation, we computed the matrix elements of both an external singlet axial-current and pseudo-scalar operator between the vacuum and a pair of external gluons in QCD with massive quarks at on-shell kinematics.
Compared to the calculations done in purely massless QCD, a consistent generation of the mass counter-terms should be done with care in Kreimer scheme.
The outcome of these checks can be concluded as our successful verification of the axial Ward identities up to $\mathcal{O}(\alpha_s^2)$ in QCD with massive quarks with the aid of the revised renormalization formula for an external (singlet) axial-current operator.

Since Kreimer scheme was known and checked to work in the several one- and two-loop calculations where the axial-current operator appears either on open fermion lines or closed fermion chains with negative superficial degree of ultraviolet divergence, the result discussed in the present work may be used to confirm that with $\gamma_5$ treated in the original Kreimer scheme and its variant, the axial-current operator needs no more additional renormalization in dimensional regularization but \textit{only} for non-anomalous amplitudes in a perturbatively renormalizable theory. %This excludes gauge theories with internal gauge axial anomalies. 
However, it would be desirable to have this statement scrutinized more stringently to completely rule out any further surprise, especially when the chiral electroweak corrections are considered.
On the other hand, based on the issues revealed through the present work, it seems that in the study of matrix elements with an \textit{external} axial anomaly, such as investigating the non-decoupling mass logarithms, calculating polarized structure functions as well as Wilsion coefficients in front of axial-current operators in some effective Lagrangian, the employment of a non-anticommuting $\gamma_5$ may be more convenient. 

\section*{Acknowledgments}

The author is grateful to Y.Q.~Ma for a chat that initiated the investigation presented in this work as well as his valuable feedback on the manuscript. 
The author is also grateful to M.~Czakon for his careful reading of the manuscript as well as many helpful comments on it. 
This work was supported by the Natural Science Foundation of China under contract No.~12205171, No.~12235008 and No.~12321005.

\appendix

\section{Extensions to an anomaly-free chiral theory in DR}
\label{append:g5pre}

The procedure described in section~\ref{sec:RoK} applies to QCD corrections to Green functions involving \textit{external} composite operators with $\g5$. 
In this case, the set of external $A_5$-vertices in eq.~\eqref{eq:FCdefG} for the definition of the trace associated with a given closed fermion chain is clear from the outset, and remains the same irrespective of the QCD-loop orders.  
Here we supply the additional ingredients needed to extend the application of the aforementioned algorithmic procedure beyond the pure QCD corrections, having in mind scattering amplitudes in a multiplicatively-renormalizable anomaly-free chiral theory, essentially the electroweak sector of the SM.
Note that due to the issue uncovered in this work, the applicability of the proposed prescription shall be, in any case, confined just to non-anomalous amplitudes in a fundamental renormalizable theory. 

\subsection{The criteria to identify the external momenta of a fermion loop}
\label{append:g5pre_IE}

The procedure in section~\ref{sec:RoK} for determining the $\oVFF$ sub-graphs of a given fermion-loop diagram $G$ requires as an input the information on the set of its external momenta, denoted as $E_Q$ in the following discussions. 
In the case of QCD corrections to Green functions with \textit{external} $\g5$-involved composite operators, the information on $E_Q$ is simply given by examining the set of external $A_5$-vertices in $G$. 
From the point of view of this procedure, the characteristic feature of the previously considered diagram $G$ 
is that each external momentum of $G$ itself equals to the difference between the momenta of certain pair of fermion propagators from the target fermion loop (which are not necessarily adjacent).

\begin{figure}[htbp]
\begin{center}
\includegraphics[scale=0.72]{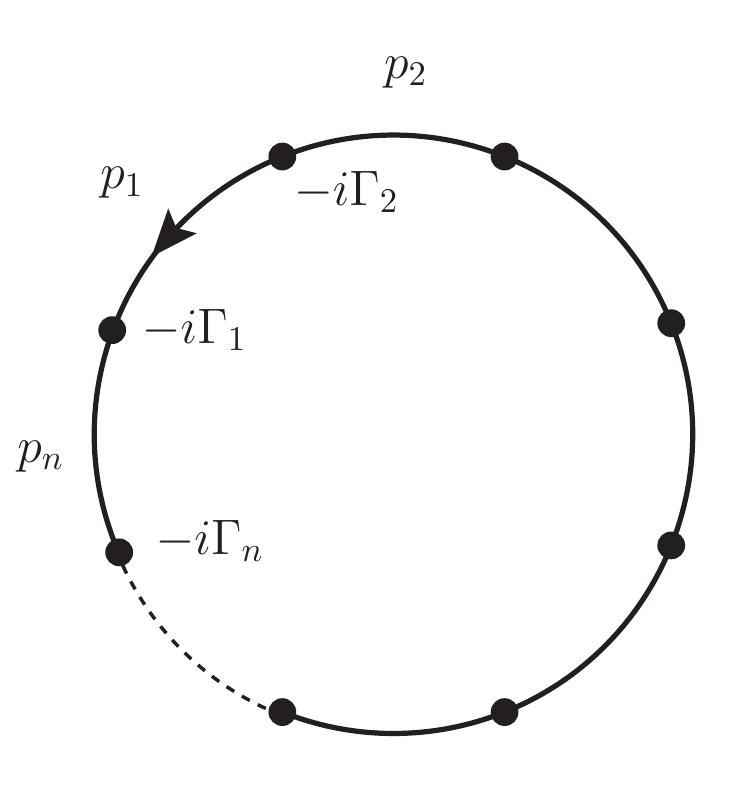}
\caption{An illustration of a fermion loop $F_G$ with $n$ fermion propagators, delimitated by black dots at which there may be some boson propagators attached, either external or internal, associated with certain vertex matrix $-i\Gamma_i$ according to the Feynman rule of the fundamental theory (which may contain $\g5$). The arrow on the solid line represents the direction of the fermion-charge flow on this fermion loop. The dashs between the two black dots represents an indefinite number of fermion propagators not drawn explicitly. The $i$-th fermion propagator's momentum  determined w.r.t.~the direction of the fermion-charge flow is denoted by $p_i$.
}
\label{fig:FermionLoop}
\end{center}
\end{figure}

Let us call the target fermion loop $F_G$, consisting of just a list of fermion propagators such as illustrated in figure~\ref{fig:FermionLoop}, 
deeply buried in a large Feynman diagram $\mathbf{G}$ with possibly electroweak loop corrections.
During the generation of $\mathbf{G}$ according to the Feynman rule in use, one must have the following knowledge on $F_G$: 
the list of fermion-propagator momenta ordered in the direction against the fermion-charge flow, which we parameterize as% 
\begin{equation}
S_p \equiv \{p_1, p_2, \cdots, p_n \}\,,    
\end{equation}%
where $p_i$ denotes the momentum value of $i$-th fermion propagator determined w.r.t.~the direction of the fermion-charge flow, e.g.~as in figure~\ref{fig:FermionLoop}. 
The ``first'' propagator with $p_1$ on this fermion loop can be arbitrarily chosen.
For the sake of later discussion, we define the following two additional sets of momenta: 
\begin{eqnarray}
S_q &\equiv& \{q_1, q_2, \cdots, q_{n-1} \} \quad  \text{with} \quad  q_{i} \equiv p_{i+1} - p_i\,, \nonumber\\    
S_Q &\equiv& \{ \cdots, p_j - p_i \cdots \} \quad  \text{for} \quad n \geq j > i \geq 1 \,.
\end{eqnarray}
$S_q$ is the set of out-going momenta through each vertices in figure~\ref{fig:FermionLoop}~(represented by black dots).
$S_Q$ amounts to be a collection of all possible $\frac{n(n-1)}{2}$ out-going momenta that can be composed from the sum of any subset of $S_q$.
Note that the total-momentum conservation does not lead to any constraint or relation among $p_i$, as one of them, say $p_1$ by default can be identified or used to parameterize the fermion-loop momentum of $F_G$ with the remaining $n-1$ (independent) $p$-momenta linearly equivalent to $S_q$. 
It can happen that each $q_i$ is associated with an internal propagator, even part of some loops, of the large Feynman diagram $\mathbf{G}$ where $F_G$ is embedded.

Having the stage set up, we are now ready to address the task, that is, how to extract the relevant information needed for defining the ($\g5$-odd) trace associated with $F_G$ embedded in $\mathbf{G}$, especially in view of the goal to maintain the most-celebrated feature of an anticommuting $\g5$ scheme (i.e.~absence of the $\g5$-specific compensation terms). 
As alluded at the beginning of this section, it is essentially the set of \textit{external} momenta $E_Q$ identified w.r.t. the sub-graph $G$ containing $F_G$, that is required as an input for the procedure in section~\ref{sec:RoK}. 
%Apparently, members of $E_Q$ must belong to $S_Q$.
From the topological point of view, what we are looking for may be identified as the \textit{minimal} cut of boson propagators of the original $\mathbf{G}$ to isolate the target fermion loop $F_G$ into a 1PI-diagram $G$ each of whose external momenta must equal to the difference between the momenta of certain pair of fermion propagators of $F_G$, plus a minimal number of remaining complementary diagrams.\footnote{These complementary diagrams can be empty if $\mathbf{G}$ fulfills the defining criteria of $G$, namely each external momentum of the 1PI $\mathbf{G}$ equals to the difference between the momenta of certain pair of fermion propagators of $F_G$. In this case, $\mathbf{G} = G$, just like those representing QCD corrections to Green functions with external composite operators considered in the bulk of this work.}

This defining criteria of $G$ containing the target $F_G$ embedded in a given $\mathbf{G}$ (generated, e.g.~in the electroweak theory) allows us to identify the set $E_Q$ of the external momenta of $G$, %the set of \textit{external} momenta identified w.r.t. the sub-graph $G$
the input information to determine the $\oVFF$ sub-graphs of $G$ for defining the trace associated with $F_G$. 
To this end, let us first introduce a subset $S_{\bar{Q}}$ of $S_{Q}$ by selecting only those $p_j - p_i$ (with $n \geq j > i \geq 1$) which does appear as the momentum flow through some boson propagator(s) in $\mathbf{G}$.
In other words, for any momentum in $S_{\bar{Q}}$ there exists at least one boson propagator in $\mathbf{G}$ with that momentum flow. 
The $E_Q$ which we are looking for is thus the smallest set made out of members of $S_{\bar{Q}}$ that fulfills simultaneously the condition that upon cutting boson propagators with momenta belonging to this subset, $F_G$ can be isolated and separated into a 1PI-subgraph $G$ with this sole subset as its external momenta.\footnote{There can be more than one propagator in $\mathbf{G}$ with a momentum in $E_Q$, corresponding to the presence of self-energy corrections to the propagator with this momentum. But this would not affect the statement here in any essential way, however, it does require care in the implementation of this procedure.} 
Apparently, one only needs to examine the subsets of $S_{\bar{Q}}$ that can satisfy the total momentum conservation\footnote{except for the cases of selecting just one single momentum, which, if valid, implies either a tadpole or a bubble graph (needing a separate treatment)}, which can be sorted according to the number of momenta therein, one by one starting from the one with the least number of momenta.

If the number of momenta in the resulting $E_Q$ is no more than two, then there can not be any contribution featuring a single Levi-Civita tensor, requiring an odd number of $\g5$ on $F_G$, simply due to the lack of sufficient momenta and/or open Lorentz indices to support such a structure.
The only exception is when the two-point fermion-loop diagram $G$ is cut open (for instance considering its value at the on-shell cut), giving rise to additional external momenta that are not inclusively integrated over.
This constitutes the only exceptional scenario that requires separate treatment.
Apparently, the non-vanishing on-shell cut of this $G$ either cuts the fermion-loop $F_G$ open %reducing it into the square/interference of certain amplitudes,
in which case the well-known open-fermion chain treatment can be applied, or leaves $F_G$ intact but with external momenta more than two (see below).

For the remaining normal cases, the number of momenta in the resulting $E_Q$ is more than two, which are assumed to be non-degenerate, i.e.~none of its proper subsets satisfies the momentum conservation.
Then, for each momentum in $E_G$, we call for the procedure in section~\ref{sec:RoK} to determine the corresponding $\oVFF$ sub-graph of $G$ regarding the target $F_G$ therein.
As long as we limit our scope to just non-anomalous amplitudes in a (multiplicatively) renormalizable fundamental theory like the SM, we believe that the $\g5$-odd trace for $F_G$ can simply be defined as an average like in eq.\eqref{eq:FCdefG}, but over each external momentum in $E_G$ irrespective of the particle species of the boson propagator with that momentum in $G$.

\subsection{Treatment of the Levi-Civita tensor in DR}
\label{append:g5pre_LV}

It is known that the Levi-Civita tensor can only be mathematically defined consistently in 4 dimensions, and mathematical inconsistency appears once one insists on the commutation in the contraction ordering for a product of multiple of them with an indifinite dimension D, due to the lack of the 4-dimensional Schouten identity.
Let us now extends the particular kind of treatment of the Levi-Civita tensors, discussed at the end of section~\ref{sec:RoK}, from application to QCD corrections (to Green functions with external composite operators) to scattering amplitudes in an anomaly-free chiral theory, e.g.~the electroweak theory.

Clearly there is one Levi-Civita tensor $\epsilon^{\mu\nu\rho\sigma}$ resulting from one $\g5$-odd trace associated with a closed fermion chain. 
This $\epsilon^{\mu\nu\rho\sigma}$ may be contracted with another $\epsilon^{\mu\nu\rho\sigma}$ either from an independent $\g5$-odd trace or external polarization-state projector, and/or some external momenta. 
As noted already at the end of section~\ref{sec:RoK}, if one wishes to manipulate $\epsilon^{\mu\nu\rho\sigma}$ with formally D($\neq 4$)-dimensional Lorentz indices for the sake of avoiding the implementation of dimensional splitting, one needs to ensure the absence of the potential compatibility issue with $\g5^2 = \hat{1}$, and also provide a rule to fix the contraction-order ambiguity among multiple $\epsilon^{\mu\nu\rho\sigma}$.
We seek to meet these requirements in the following way.
\begin{itemize}
\item 
The relation $\g5^2 = \hat{1}$ is applied always for any pair of $\g5$ on the same fermion chain, and 
mathematical identities involving $\g5$ in Dirac algebra specific to 4-dimensions, such as the Fietz identities and (inverse) Chisholm identity, should not be applied in general.
A pair of $\epsilon^{\mu\nu\rho\sigma}$ to be contracted must come from $\g5$-odd traces associated with two \textit{different} fermion loops and/or external projectors associated with vector-boson polarization states.

\item 
Denote the contraction of a general product of Levi-Civita tensors in the problem by 
\begin{equation}
\label{eq:LVTC}
\big[\epsilon_{i_1}\, \, \, \cdots\, \epsilon_{i_{N_i}}\big]\,  
\big[\epsilon_{e_1}\, \, \cdots\, \epsilon_{e_{N_e}}\big] 
\end{equation}
where the Lorentz indices are suppressed for the sake of concise notations, and the $[~]$ wrapping around a product of Levi-Civita tensors indicates the contraction operation.
Furthermore, we have introduced two sets of subscripts for Levi-Civita tensors:
those labeled by $e_k$ for $k = 1,\cdots, N_e$ are from the external projectors associated with vector-boson polarization states; 
those labeled by $i_k$ for $k = 1,\cdots, N_i$ are from the independent $\g5$-odd fermion loops.

To eliminate the potential contraction-order ambiguity in eq.\eqref{eq:LVTC}, we proceed as follows.
\begin{itemize} 
\item 
For $\big[\epsilon_{e_1}\, \, \cdots\, \epsilon_{e_{N_e}}\big]$ in eq.\eqref{eq:LVTC} which appears as an overall factor for the quantities in question, simply mark each of them and fix an arbitrary but definite contraction-order, which shall be adopted consistently in the calculation. %(which can also be left undone to the very end).

\item 
For $\big[\epsilon_{i_1}\, \, \, \cdots\, \epsilon_{i_{N_i}}\big]$ in eq.\eqref{eq:LVTC}, take the symmetric average over all possible pairings.\footnote{We thank Prof.~Yan-Qing~Ma for pointing this out.} 
For instance in the case of $N_i = 4$, we define 
\begin{eqnarray*}
\big[\epsilon_1\, \epsilon_2\, \epsilon_3\, \epsilon_4\big] 
\equiv 
\frac{1}{3} 
\big( 
\big[ \epsilon_1\, \epsilon_2 \big] 
\big[ \epsilon_3\, \epsilon_4 \big]
\,+\, 
\big[ \epsilon_1\, \epsilon_3 \big] 
\big[ \epsilon_2\, \epsilon_4 \big]
\,+\, 
\big[ \epsilon_1\, \epsilon_4 \big] 
\big[ \epsilon_2\, \epsilon_3 \big]
\big)\,.
\end{eqnarray*}
The number of the possible pairings to be averaged over in $\big[\epsilon_{i_1}\, \, \, \cdots\, \epsilon_{i_{N_i}}\big]$ are 
\begin{equation*}
\begin{cases}
& \frac{N_i !}{(N_i/2)!\,2^{N_i/2}}  \quad \text{if $N_i$ is even}\,, \\
& N_i\,\frac{(N_i-1)!}{((N_i-1)/2)!\,2^{(N_i-1)/2}}  \quad \text{if $N_i$ is odd}\,, 
\end{cases}
\end{equation*}

\item 
The results derived respectively for $\big[\epsilon_{i_1}\, \, \, \cdots\, \epsilon_{i_{N_i}}\big]$ and $\big[\epsilon_{e_1}\, \, \, \cdots\, \epsilon_{e_{N_e}}\big]$ are then multiplied together as in eq.\eqref{eq:LVTC}.
\end{itemize}

\item 
The contraction between any single pair of $\epsilon^{\mu\nu\rho\sigma}$ involved above is done according to eq.\eqref{eq:LeviCivitaContRule}, with the resulting spacetime-metric tensor $g_{\mu\nu}$ set $D$-dimensional. 

\end{itemize}

Applying the whole procedure described in this Appendix to a given set of Feynman loop diagrams \textit{defines} the corresponding (non-anomalous) amplitudes with (odd) $\g5$ on the closed fermion chains in our prescription.
The prescription is formulated such that the involved algorithmic procedures can be readily implemented in the computer-algebra tools currently available on the market for generating Feynman diagrams and performing Dirac algebra, without reference to the notion of non-cyclic trace. 
We emphasize again that due to the issue uncovered in this work, the applicability of the proposed prescription shall be, in any case, confined to non-anomalous amplitudes in a fundamental renormalizable theory, like the electroweak sector of the SM. 
We note briefly that the appearance of \textit{intermediate} IR divergence (such as those associated with final-state radiations) in the on-shell cuts of loop amplitudes should not pose any conceptually new difficulties. 
The guiding principle for a proper treatment is that at least for a minimal set of contributing pieces where the intermediate IR divergence are supposed to cancel between each other resulting an IR-finite sum, the very same $\g5$ treatment shall be applied consistently in deriving the respective bare expressions, i.e.~independent of the specific configurations of the on-shell cuts distinguishing these individual pieces.

It is interesting to note that the 1PI amplitudes with non-negative degree of superficial UV divergence in the SM with two $\g5$-odd fermion loops only start from three-loop order (see, e.g.~\cite{Mihaila:2012pz}), which are actually UV-finite as a whole.
We thus expect that the treatment described above shall work for the SM at least to three-loop order with the usual formally-proved multiplicative renormalization structure that are typically presented in literature without worrying much about the $\g5$ issue.

We conclude this Appendix by the following comment:
the validity of this particular prescription in applications to the full SM to three-loop order is currently still a conjecture. 
Needless to say, this expectation needs to be verified, or refuted, by explicit calculations involving multi-loop electroweak corrections, which are left for future works.

\bibliography{acg5} 
\bibliographystyle{utphysM}
% ********** Ending **********
\end{document}